\documentclass{aa}  

\usepackage{graphicx}
\usepackage{txfonts}
\usepackage{soul}
\usepackage{float}
\usepackage[dvipsnames]{xcolor}
\usepackage[colorlinks=true, citecolor=blue]{hyperref}
\usepackage{placeins}
\usepackage{multirow}
\usepackage{pbox}
\usepackage{graphicx}	
\usepackage{amsmath}	
\usepackage{amssymb}	
\usepackage{multirow}
\usepackage{gensymb}
\usepackage{booktabs}
\usepackage{pdflscape}
\usepackage[flushleft]{threeparttable}
\usepackage{color}
\usepackage{nidanfloat}

\newcommand{\fgas}{$f_{\rm gas}$}

\newcommand{\gdr}{GDR}

\newcommand{\sfrunit}{$\rm M_{\odot}\,yr^{-1}$}

\newcommand{\s}{{\it Spitzer}}
\newcommand{\h}{{\it Herschel}}
\newcommand{\msol}{$\rm M_{\odot}$}
\newcommand{\msoll}{$M_{\rm\odot}$}
\newcommand{\md}{$M_{\rm dust}$}

\newcommand{\lsol}{L$_{\odot}$}
\newcommand{\ms}{$M_{\ast}$}

\newcommand{\td}{$T_{\rm d}$}

\newcommand{\mgas}{$M_{\rm gas}$}
\newcommand{\depl}{$\tau_{\rm dep}$}

\newcommand{\um}{$\langle U \rangle$}

\newcommand{\myr}{$M_{\odot}$ yr$^{-1}$}
\newcommand{\msun}{$M_{\odot}$}
\newcommand{\mstar}{$M_{\ast}$}
\newcommand{\mdust}{$M_{\rm dust}$}

\newcommand{\lir}{$L_{\rm IR}$}

\newcommand{\lsun}{$L_{\odot}$}

\newcommand{\tdust}{$T_{\rm dust}$}

\newcommand{\tdep}{$t_{\rm dep}$}

\begin{document}

   \title{The interstellar medium of quiescent galaxies and its evolution with time}

   \author{Georgios E. Magdis\inst{1,2,3}
    \and Raphael Gobat \inst{4}
    \and Francesco Valentino \inst{1,3}
    \and Emanuele Daddi \inst{5}
    \and Anita Zanella \inst{6}
    \and Vasily Kokorev \inst{1,3}
    \and Sune Toft \inst{1,3}
    \and Shuowen Jin \inst{7,8}
    \and Katherine E. Whitaker \inst{9,1}
     }

   \institute{Cosmic Dawn Center (DAWN), Copenhagen, Denmark\\
              \email{geoma@space-dtu.dk}
         \and DTU-Space, Technical University of Denmark, Elektrovej 327, DK-2800 Kgs. Lyngby, Denmark
         \and Niels Bohr Institute, University of  Copenhagen, Lyngbyvej 2, DK-2100 Copenhagen \O
         \and Instituto de Fisica, Pontificia Universidad Catolica de Valparaiso, Casilla 4059, Valparaiso, Chile          
         \and CEA, IRFU, DAp, AIM, Université Paris-Saclay, Universit\'e Paris Diderot, Sorbonne Paris Cité, CNRS, F-91191 Gif-sur-Yvette, France
         \and Istituto Nazionale di Astrofisica, Vicolo dell'Osservatorio 5, 35122 Padova, Italy 
         \and Instituto de Astrofísica de Canarias (IAC), E-38205 La Laguna, Tenerife, Spain 
         \and Universidad de La Laguna, Dpto. Astrofísica, E-38206 La Laguna, Tenerife, Spain
         \and Department of Astronomy, University of Massachusetts, Amherst, MA 01003, USA
             }

   \date{Received --; accepted --}

 
  \abstract{ We characterise the basic far-IR properties and the gas mass fraction of massive ($\langle$ log(\ms/\msoll) $\rangle \approx 11.0$) quiescent 
   galaxies (QGs) and explore how these evolve from $z = 2.0$ to the present day.  We use robust, multi-wavelength (mid- to far-IR and sub-millimetre  to radio) stacking ensembles of homogeneously selected and mass complete samples of log(\ms/\msoll) $\gtrsim 10.8$ QGs.  We find that the dust to stellar mass ratio (\md/\ms) rises 
   steeply as a function of redshift up to $z \sim 1.0$ and then remains flat 
   at least out to $z = 2.0$.  Using \md\ as a proxy of gas mass 
   (\mgas), we find a similar trend for the evolution of the gas mass 
   fraction (\fgas), with $z > 1.0$ QGs having \fgas\ $\approx 7.0\%$ (for solar metallicity). This \fgas\ is three to ten times lower than that of normal star-forming galaxies (SFGs) at their corresponding redshift but $\gtrsim$3 and $\gtrsim$10 times larger  compared to that of $z = 0.5$ and local QGs. Furthermore, the inferred gas depletion time scales are comparable to those of local SFGs and 
    systematically longer than those of main sequence galaxies at their
    corresponding redshifts. Our analysis also reveals that the average dust temperature (\td) of massive QGs  
    remains roughly constant ($\langle$ \td $\rangle = 
    21.0 \pm 2.0$\,K) at least out to $z \approx 2.0$ and is substantially colder ($\Delta$\td\ $\approx$ 10\,K) compared to that of SFGs. This motivated us to construct and release a redshift-invariant template IR SED, that we used to make predictions for ALMA observations and to explore systematic effects in the \mgas\ estimates of massive, high$-z$ QGs. Finally, we discuss how a simple model that considers progenitor bias can effectively reproduce the observed evolution of \md/\ms\ and \fgas. Our results  indicate universal initial interstellar medium conditions for quenched galaxies and a large degree of uniformity in their internal processes across cosmic time.
    }
    

   \maketitle
%

\section{Introduction}

The existence of massive, passively evolving galaxies at various cosmic epochs is now well established thanks to observational efforts that span more than two decades, from first discoveries  \citep[e.g.][]{Daddi00, daddi05, Cimatti_04,  Toft05, toft07, Kriek_06} to the latest spectroscopic confirmations  \citep[e.g.][]{toft12,toft17, Gobat12,Whitaker13, Glazebrook17,Schreiber18, Tanaka19,V20,DEugenio20,Stockmann20}. This population of early type galaxies (ETGs), or more generally of quiescent galaxies (QGs){\footnote{in this paper we use the terms ETG and QG interchangeably.}}, experiences low (or negligible) levels of star formation, resides well below the  main sequence (MS) of star formation \citep[e.g.][]{Noeske07,daddi07,magdis10, Speagle14,schreiber15_ms} at their corresponding redshift, and is characterised by red colours, indicative of old and evolved stellar populations. With a large fraction of  $z \geq$ 2 massive galaxies (\ms\ $> 10^{11}$\,\msol) experiencing a downfall in their star formation activity \citep[e.g.][]{Whitaker_2012,Davidzon17}, the abundance of QGs progressively increases at later times. At the same time, recent studies have attested that a fraction of them are already in place by $z\sim4$ \citep[e.g.][]{Glazebrook17,Schreiber18, Tanaka19,V20}. 

While the mere detection of passive and massive galaxies out to $z\sim 3-4$  poses in itself challenges in our theories of galaxy evolution, their existence is perplexing at a more basic level: The very mechanisms responsible for triggering and/or sustaining the cessation of star formation at any redshift remain an unsolved puzzle (for a recent overview, see \citealt{Man18}). Several competing scenarios that attempt to explain quenching have been put forward, promoting mechanisms that prevent gas from cooling (halo quenching, active galactic nuclei and stellar feedback, strangulation, e.g. \citealt{Cattaneo06, Peng15,Henriques15}), as well as gas expulsion via feedback (outflows, e.g.  \citealt{Dimatteo05,Hopkins06}) and gas stabilisation (morphological quenching, e.g. \citealt{martig09}). While these plausible quenching mechanisms invoke different processes at different physical scales (from dark matter halos to nuclear activity),  it is  evident that they all have some common parameters at their core: the  gas mass budget, the physical conditions, and the dynamical state of the interstellar medium (ISM). This  should not come as a surprise. Since gas and dust are the agents of star formation, it quite naturally follows that they should also be at the heart  of its cessation, or in other words, for quenching. Consequently, progress in the field necessitates a careful study of the ISM of QGs, very similar to what has already been achieved for star-forming galaxies (SFGs). 
 
To this end, a large volume of studies have  scrutinised the ISM properties of local QGs.  By exploring their far-IR (FIR) and millimtetre (mm)  dust continuum emission and/or targeting atomic and molecular lines (e.g. low transition CO lines), these studies have revealed low, but not negligible, amounts of dust \citep[e.g.][]{Gomez10, Rowlands12, Smith12, Agius13, Werner14, Boselli14, Lianou16}  and gas \citep[e.g.][]{Saintonge11a1, Saintonge11a2, Young11, Cappellari13, Davis14, Boselli14, French15, Alatalo16} in local QGs and overall FIR characteristics that are in contrast to those of local SFGs. In particular, the dust mass fraction (\md/\ms) of local QGs is found to vary  between 10$^{-5}$ and 10$^{-4}$, and their molecular gas fraction (\fgas = \mgas/\ms) from 0.3  to 1\%, with both quantities being $\sim 2$ orders of magnitude lower than those of normal galaxies in the local Universe. Similarly, their typical `luminosity-weighted' dust temperatures (\td) are $\sim$20\,K \citep[e.g.][]{Smith12}, on average 3-5\,K colder than those of normal SFGs at $z = 0$.
 
On the other hand, the exploration of the properties of the ISM in QGs  beyond the local Universe is still in its infancy, in part due to their strenuous  selection (and spectroscopic confirmation), but more importantly due to their faint FIR nature. In spite of these challenges though,  some `brave' attempts to measure the  \mgas\ of distant QGs and post-starburst (pSB) galaxies  have already been carried out \citep[e.g.][]{Sargent15,Suess17,Rudnick17,Hayashi18,Spilker18,Bezanson19,Williams_20}.  These studies of small and inhomogeneous (in terms of selection, and physical properties, e.g. \ms) samples of distant of QGs, have so far provided only loose (if any) constrains in their gas mass budget and its evolution with time, placing their \fgas\ anywhere between 15\% and $<$3\%. The diverse  \fgas\ measurements emerging from various studies demonstrate the need for a systematic study of the ISM for a carefully and uniformly selected sample of QGs in various redshifts bins. 

At the same time, the shape of the FIR spectral energy distributions (SEDs) of distant QGs remains even less explored. Indeed, while reasonable \mgas\ and \md\ estimates can be achieved with single line (e.g. CO) or single-band dust continuum observations, a full characterisation of their IR SEDs requires multi-wavelength photometry probing both the Wien part and the Rayleigh-Jeans (R-J) tail, a prohibitively expensive task in terms of observing time, even for the limited number of available QGs that benefit from lensing magnification. As such, any progress on this front can only be achieved though stacking  \citep[e.g.][G18 hereafter]{Viero13, Man16, Gobat_2018}.  

Finally, it is worth recalling that the detailed study of the FIR properties of representative and complete samples of SFGs at various cosmic epochs has advanced our understanding of galaxy evolution through scaling relations and evolutionary tracks that have been used to guide simulations and theoretical models \citep[e.g.][]{Dekel09, Popping14, Narayanan15, Lagos15, Popping17,Dave17,Dave20}. Far-IR studies have revealed that the gas and dust mass fraction of SFGs have sharply declined over time, with SFGs having an ISM mass budget larger by a factor of $\sim$10 at $z \approx 2$  compared to that of SFGs of similar \ms\ in the local Universe \citep[e.g.][]{daddi_2008, daddi_2010, geach11, magdis_2012,magdis17, Tacconi18, Liu19}. At the same time high$-z$ SFGs, despite their higher star formation rates (SFRs),  are characterised by shorter gas depletion time scales (\depl) indicative of a higher star formation efficiency at earlier times \citep[e.g.][]{Tacconi18, Liu19}.  Similarly,  the \td\ of SFGs also increases with redshift  \citep[e.g.][]{magdis_2012,magdis17, Magnelli14, schreiber_2018_dust, Cortzen20}, mirroring an evolution in their star formation activity,  star formation surface density, and the strength of their radiation field. As a  final example, we refer to  the construction of representative FIR SEDs of SFGs at various redshifts \citep[e.g.][]{ elbaz11, magdis_2012, schreiber_2018_dust}, which have greatly facilitated models, predictions, and extrapolations.

These successes motivated us to undertake a similar effort, but this time for the population that consists of the ending point of the evolutionary path of SFGs, in other words, for QGs. In this work we aim to characterise the basic FIR properties (\lir, \td, \md, \mgas, \fgas, \depl), and trace their evolution with time, for a robust and controlled sample of QGs in various redshift bins. For this task we employed a multi-wavelength, mid- to far-IR and millimetre to radio stacking analysis of homogeneously selected and mass complete (log(\ms/\msoll) $\gtrsim 10.8$) QGs samples in the  $z = 0.3 - 1.5$ redshift range, drawn from the COSMOS field. To extend the redshift coverage of our work we further complement our analysis with the $1.5 < z < 2.0$ stacked sample of QGs presented in \citet{Gobat_2018} and which is constructed using the same selection criteria as those employed in our study.

The paper is organised as follows: in Section 2 we present the  selection criteria  for our samples of QGs in various redshift bins. In Section 3 we present the stacking methodology and the subsequent photometry.  In Section 4 we present a detailed analysis of the FIR properties of our stacked ensembles and in Section 5 we explore their evolution with time. In Section 6 we discuss the implications of our findings and provide an outlook for future IR studies of distant QGs with (sub-)millimetre facilities. Finally, in Section 7 we summarise the main findings of our work. Throughout the paper we adopt a \citet{salpeter} initial mass function (IMF) and  H$_{0}$ = 70\,km\,s$^{-1}$ Mpc$^{-1}$, $\rm \Omega_{\rm M}$ = 0.3, and $\rm \Omega_{\rm \Lambda}$ = 0.7.

\section{Selection}\label{sec:selection}
\begin{figure*}
\includegraphics[width=0.45\textwidth]{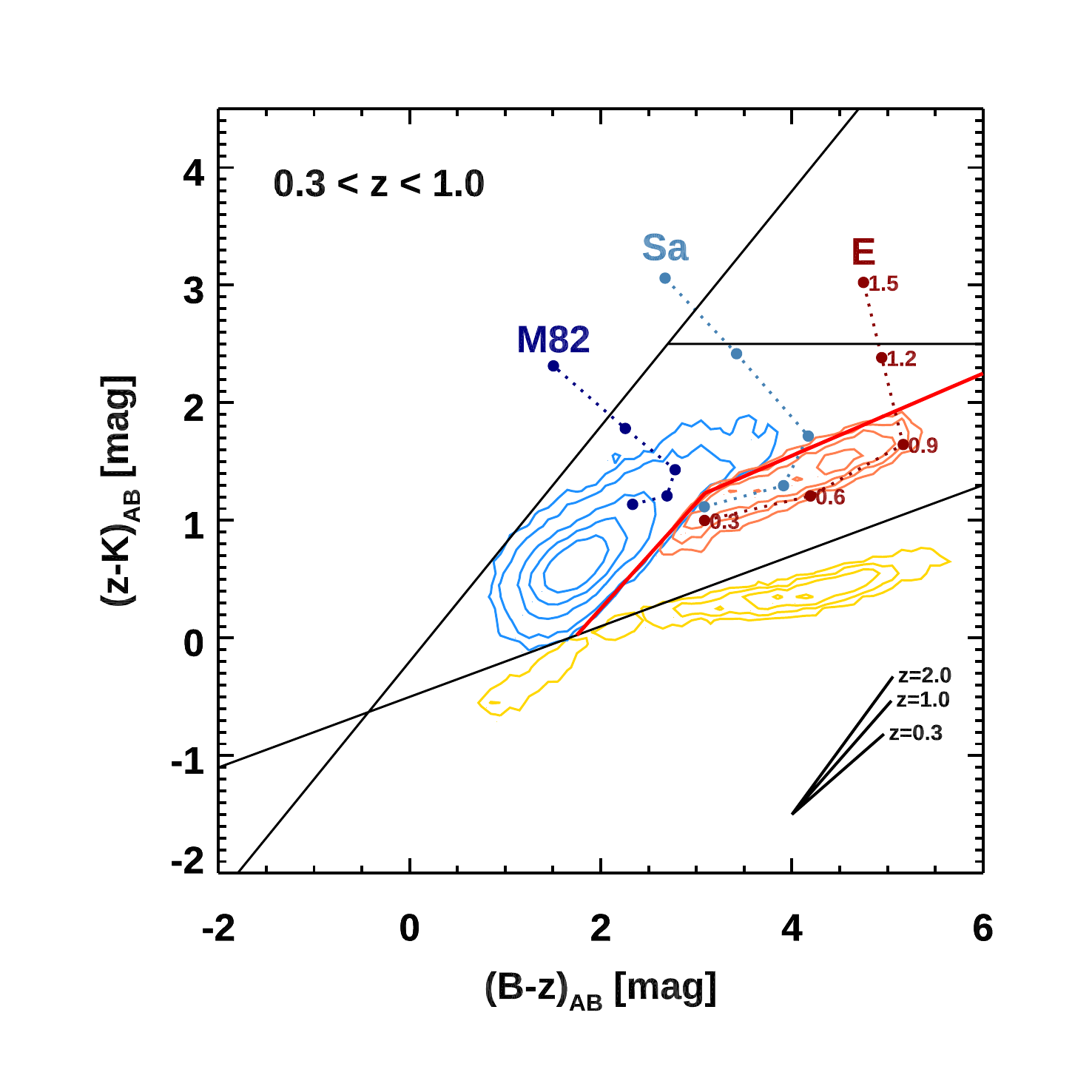}
\includegraphics[width=0.45\textwidth]{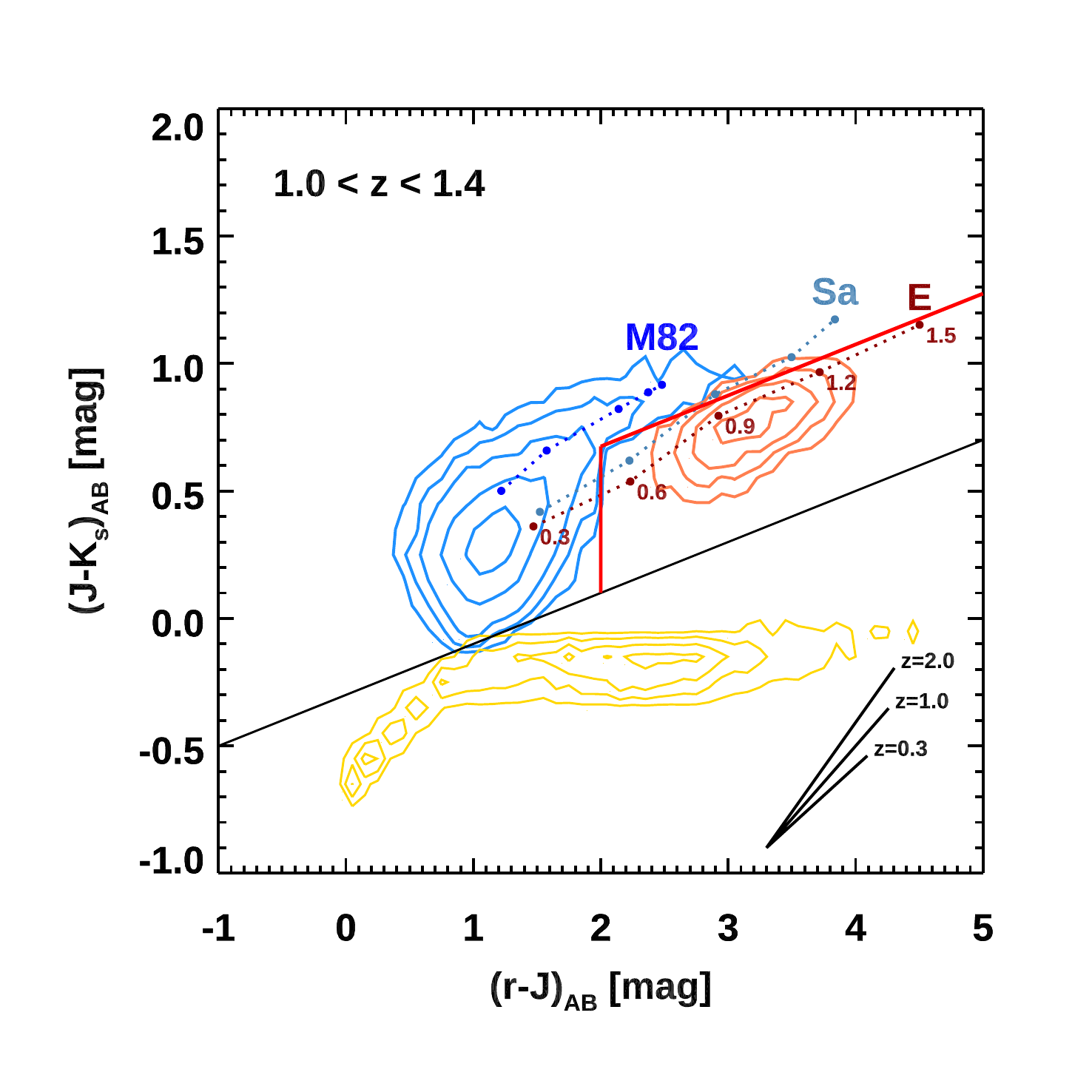}
\caption{\textbf{Redshift ranges for an efficient use of the \textit{BzK} and
  \textit{rJK} colour diagrams}. \textit{Left:} \textit{BzK} diagram for the selection
  of $0.3 \leq z < 1.0$ QGs. The yellow, blue, and red contours represent the
  loci of stars, SFGs, and QGs at $0.3 \leq  z < 1.0$ selected by the $NUVrJ$+$BzK$ selection. 
  Sources with a $3\sigma$ detection at $24\,\mu$m in \cite{jin_2018} have been excluded. The black lines
  represent the original cuts by \citet{daddi_2004} to select
  galaxies at $1.4\lesssim z \lesssim 2.5$. The red lines show our
  cut at low redshifts (Eq. \ref{eq:bzk_cut}). For reference, the tracks show the 
  synthetic colours for the
  13\,Gyr elliptical (E), early-type spirals
  (Sa), and M82-like templates by \citet{polletta_2007}, as marked in the labels. The
  colours are computed for $0.3 < z < 1.5$ in step of $\Delta z = 0.3$. No evolutionary effects are included. The black
  segments show the reddening vectors at $z = 0.3$, $1$, and
  $2$. \textit{Right:} \textit{rJK} diagram for the selection of $1.0\leq  z < 1.4$ QGs. The yellow, blue, and red contours represent the
  loci of stars, SFGs, and QGs at $1.0\leq  z < 1.4$ selected by the $NUVrJ$+$rJK$ selection. The remaining lines are coded as in the left panel.}
\label{fig:polletta07}
\end{figure*}

We selected QGs in the COSMOS field \citep{scoville_2007} based on the latest public photometric catalogue by \cite{laigle_2016}. First, we removed
stars (i.e. \texttt{TYPE=1} flag) and set a magnitude cut at $K_{\rm s} = 24.5$\,mag, the $3\sigma$ depth in $2"$ diameter apertures in the `deep' UltraVista stripes \citep{laigle_2016}.
We further excluded X-ray \textit{Chandra}-detected sources from the latest catalogs by
\cite{civano_2016} and \cite{marchesi_2016}, thus removing active galactic nuclei and strongly star-forming objects. Then, following the approach in G18, we separated  red QGs from blue
star-forming sources by combining the rest-frame \textit{NUVrJ} criterion
  \citep{arnouts_2007, ilbert_2013} with the observed 
\textit{BzK} \citep{daddi_2004} and (a custom) \textit{rJK} diagrams.\\

\textbf{The \textit{BzK} criterion.}
The \textit{BzK} diagram was originally designed to select galaxies at
$1.4\lesssim z\lesssim 2.5$, but it can also be successfully used at lower
redshift \citep{bielby_2012a}. We computed the \textit{B-z},
\textit{z-K} colours using magnitudes in 2'' diameters apertures, following
\cite{mccracken_2010, mccracken_2012}. 

We applied a $+0.05$\,mag
correction to the \textit{z-K} colour to better match the stellar
locus from the \cite{lejeune_1997} stellar models reported in the
original selection from \cite{daddi_2004}. Red QGs at $0.3\leq z
< 1.0$ fall in
the portion of the \textit{B-z}, \textit{z-K} plane delimited by:  
\begin{equation}
  \label{eq:bzk_cut}
  \begin{cases}
    z-K > 0.3\times(B-z) - 0.5\\
    z-K \leq 0.35\times(B-z) + 0.15\\
    z-K \leq 0.9\times(B-z)-1.55~ .\\
  \end{cases}
\end{equation}
We selected galaxies detected at $5\sigma$ in the $K_{\rm s}$ and $z^+$
bands, and  retained sources with lower limits on the $B-z$ colour
compatible with the cuts.\\

\textbf{The \textit{rJK} criterion.}
We adapted the principle of the \textit{BzK} diagram (i.e. two bands
bluer and one band redder than the Balmer/4000\,\AA\ jumps at the
redshift of reference) to classify galaxies in the range $1.0\leq z
<1.4$. We computed the \textit{r-J}, \textit{J-K} colours using
magnitudes measured in 2'' diameter apertures. We selected
red quiescent objects as:
\begin{equation}
  \label{eq:rjk_cut}
  \begin{cases}
    J-K > 0.2\times(r-J) - 0.3\\
    J-K < 0.2\times(r-J) + 0.275\\
    r-J > 2.0~ .\\
  \end{cases}
\end{equation}
We selected sources detected at $5\sigma$ in the $K_{\rm s}$ and $J$
band, and  retained objects with lower limits on the $r-J$ colour
compatible with the cuts.\\

We show the observed colour diagrams and the loci occupied by red QGs at $0.3 \leq z \leq 1.4$ in Figure \ref{fig:polletta07}. For guidance, we show
the synthetic colours for a series of templates of different galaxy types from from \cite{polletta_2007} (see
\citealt{bielby_2012a} for a similar exercise with
\citealt{bruzual_2003} stellar population models). We computed the
colours at different redshifts, but without including  evolution effects. The locus of 13 Gyr-old QGs falls entirely within the region occupied by $z \lesssim 1$ QGs in the $BzK$ diagram and $z \lesssim 1.4$ QGs in the $rJK$ one. Templates typical of strongly star-forming galaxies lie
entirely outside the quiescent region in both diagrams.
Early-type spirals initially fall in the star-forming
portions of the colour diagrams, later entering the quiescent region.\\
\begin{figure*}
    \includegraphics[width=\textwidth]{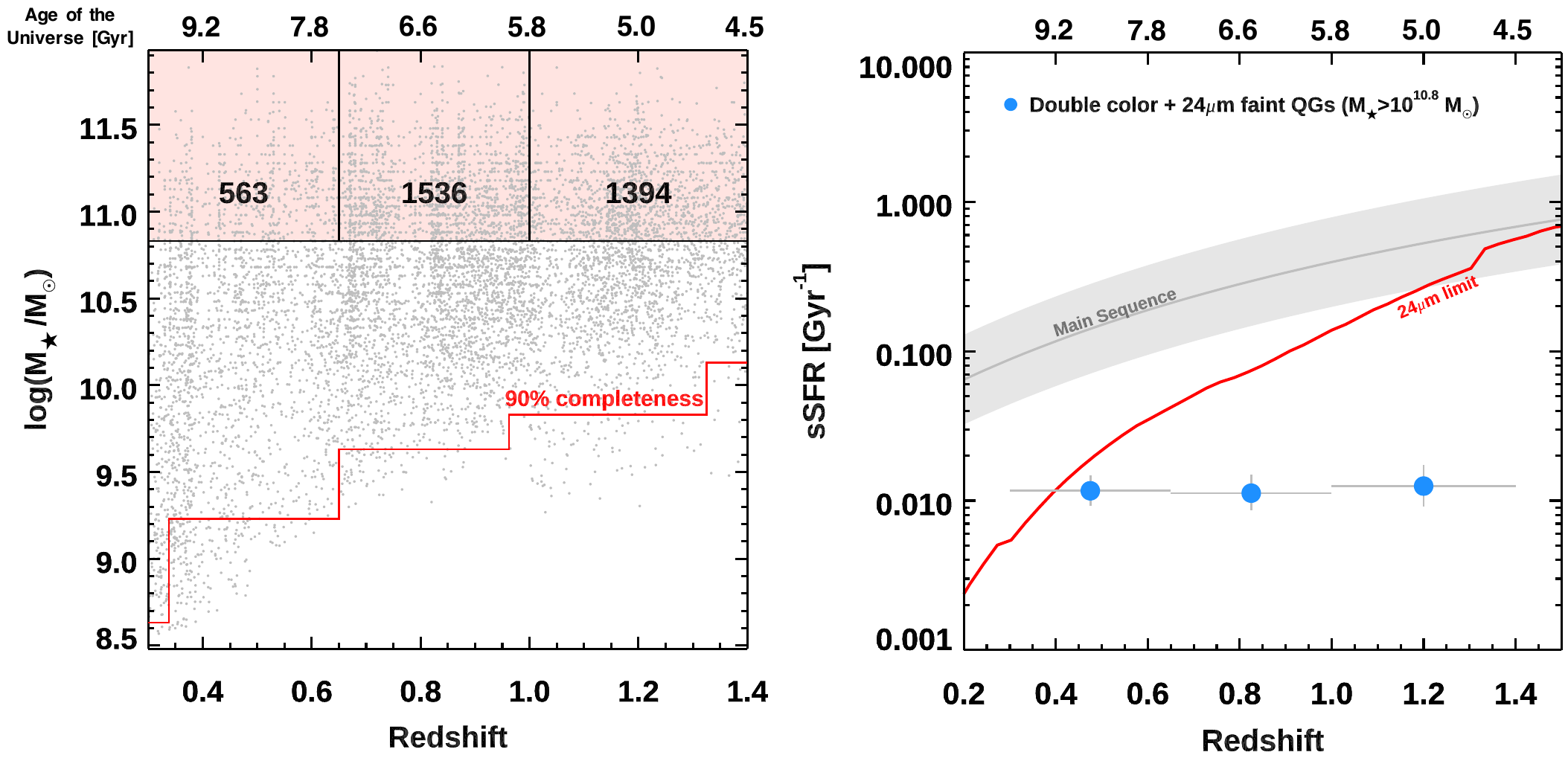}
    \caption{\textbf{Selection and binning}. \textit{Left:} Grey points showing the final sample of double-colour selected and 24 $\mu$m faint QGs in the redshift vs stellar mass plane. The rose area and the black grid mark the limits of the bins used to group
    galaxies for stacking. The number of stacked sources per bin are displayed. The red line indicates the 90\% stellar mass completeness limit for QGs in the `deep' UltraVISTA sample \citep{laigle_2016}, adjusted for the different IMF.
    \textit{Right:} Blue points showing the position of the stacked massive QGs ($M_\ast>10^{10.8}\, M_\odot$) in the redshift vs sSFR(UV) plane. The error bars on the redshift indicate the width of the bins. Only the statistical uncertainties on sSFR(UV) are shown on the y-axis. The solid red line indicates the $3\sigma$ detection limit for \textit{Spitzer}/MIPS at $24\,\mu$m ($\sigma = 15\,\mu$Jy, \citealt{jin_2018}) converted to SFR using the MS templates from \cite{magdis_2012} and assuming  $M_\ast = 10^{11.2}\, M_\odot$, matching the median stellar mass of our sample in each redshift bin. The grey shaded area and the grey line indicate the location of the MS and its 0.3\,dex scatter, computed for a galaxy with $M_\ast = 10^{11.2}\, M_\odot$.}
  \label{fig:24um_detections}
  \label{fig:binning}
\end{figure*}

The addition of the \textit{BzK} and \textit{rJK} criteria removes more efficiently the red dusty galaxies contaminating the sample of quiescent objects selected with the \textit{NUVrJ} diagram only. Sources without a constraint on the observed colours occupy regions of the \textit{NUVrJ}
diagram with bluer \textit{r-J} and redder \textit{NUV-r} colours, 
partially overlapping with the final quiescent sample only below $z<1$. On the other hand, galaxies excluded by the
\textit{BzK} and \textit{rJK} criteria lie close to the line
separating the star-forming population. A 2D Kolmogorov-Smirnov test \citep{fasano_1987} suggests that both the unconstrained and excluded galaxies are drawn from a parent
distribution different from the one of the final quiescent sample.\\

The validity of the double-colour selection is further supported by the lower fraction of \textit{Spitzer}/MIPS
$24\,\mu$m $3\sigma$ detected red galaxies when using
\textit{NUVrJ}+\textit{BzK} and \textit{NUVrJ}+\textit{rJK} colour
selections ($\sim15$\%), rather than a simple \textit{NUVrJ} diagram ($\sim25$\%). For the comparison we use the latest $24\,\mu$m catalogue by \citet{jin_2018}, reaching an rms of $\sim15\,\mu$Jy. Figure
\ref{fig:24um_detections} shows that we can
trace SFRs down to several times below the MS parameterised
as in \citet{schreiber_2015}. 
To compute the SFR limit in \citet{jin_2018}, we
rescaled the redshift-dependent, MS templates by \cite{magdis_2012} to match
the observed $24\,\mu$m depth and then integrated over the wavelength
range $8-1000\,\mu$m to derive the total infrared luminosity
(\lir). We smoothed over the width of the \textit{Spitzer}/MIPS
24\,$\mu$m filter to avoid the strong impact of the PAH
features. 

We excluded all the
$24\,\mu$m detected sources in the \citet{jin_2018} catalogue from our sample of QGs. We remark that the sample in the highest redshift bin could still potentially be contaminated by a fraction of galaxies closer to the MS, given the looser constraints on the IR emission. However, most of the contamination in the quiescent sample is due to dusty starbursts (or transitioning galaxies), rather than blue objects on the MS with undetectable IR emission. Indeed, absent IR detections, the median sSFR(UV) of the final stacked samples of massive quiescent objects is constant over the redshift interval under consideration. Therefore, the median location with respect to the MS decreases with redshift, given the steady increase in the normalisation of the former (Figure \ref{fig:binning}).\\

For the purpose of stacking, we further excluded the 1\% of the sample that lay too close to the edges of the IR image mosaics. Finally, we binned our sample by redshift, as shown in Figure \ref{fig:binning}. We split the galaxies selected with the $rJK$ colour at $z = 0.65$, the mid-point of the redshift interval over which this criterion is effective. In this work we focus on a complete sample of massive galaxies at $0.3 < z < 1.5$ (log(\ms/\msoll) $\geq 10.83)${\footnote{This corresponds to a log(\ms/\msoll) $\geq 10.6$ mass cut in the original \citet{laigle_2016} catalogue, which is built assuming a Chabrier IMF.}}, matching the median stellar mass of similarly selected objects at $\langle z \rangle = 1.8 $ from G18. This allows us to consistently trace the evolution of the dust content over 10\,Gyr of cosmic evolution. We report the properties of the stacked sample in Table~\ref{tab:properties}. 

\section{Source stacking and extraction}\label{sec:method}
Our stacking and source extraction methodology broadly follows that described in G18, namely:
For each redshift bin, we extracted cutouts, centered on each galaxy, from publicly available FIR imaging of the COSMOS field at 24\,$\mu$m \citep{lefloch_2009}, 100\,$\mu$m, and 160\,$\mu$m \citep{lutz_2011}, 250\,$\mu$m, 350\,$\mu$m, and 500\,$\mu$m \citep{oliver_2012}, 850\,$\mu$m \citep{geach_2017}, 1.1\,mm \citep{aretxaga_2011}, 10\,cm \citep{smolcic_2017}, and 20\,cm \citep{schinnerer_2010}. For each band, we then combined the cutouts to create a median image, ignoring masked (e.g. zero-coverage) regions, and estimated the variance of each pixel through bootstrap resampling of the data with half the sample size. We note that these variance cutouts do not show any appreciable structure (i.e. they are flat), suggesting that there is little to no contribution from clustering to the noise \citep[e.g.][]{Bethermin_12}.

As in G18, we model the emission in each median image with a combination of  three components: point-like emission from the central QG, an autocorrelation term due to source overlap within the sample, and extended emission from SFGs associated with the central QGs. This includes both true SF satellites of the QGs and the two-halo term arising from clustering with more massive SFGs. For simplicity, we hereafter refer to it as the `satellite halo'. Finally, we also include a flat background term. As the point-source and autocorrelation terms arise from the same galaxy population their amplitudes are tied and we determine their relative scaling beforehand. To model the combined central point-like emission and autocorrelation signal, we first created simulated images of the COSMOS field for each redshift bin, consisting of a blank map populated by model or empirical beams (depending on the data) located at the position of each QG and normalised to the same unitary flux. This latter approximation was used because the scatter of the FIR flux of our QGs is unknown by construction of the sample. Consequently, we do not associate an uncertainty with the shape of the autocorrelation signal. The simulated sources were then extracted using the same procedure as described above. This yields median cutouts that are essentially indistinguishable from point-sources in the case of higher-resolution data (e.g. at 24\,$\mu$m) but deviate substantially from it when the beam is large, due to the blending of sources. In particular, the SPIRE (250\,$\mu$m, 350\,$\mu$m, and 500\,$\mu$m) `point' stacks have therefore extended wing-like emission and higher peak values than a simple beam, while the 850\,$\mu$m and 1.1\,mm point stacks show an opposite effect, with lower peak values, arising from the deep negative rings in their PSFs. This additional step, which accounts for the intrinsic clustering of QGs, is  necessary here since we did not eliminate projected pairs during selection, as in G18. Indeed, excluding from the sample all objects that  partially overlap in the lowest-resolution maps (SPIRE) would more than halve it\footnote{As a sanity check we have validated that an analysis excluding the pairs yields consistent results albeit with significantly higher uncertainties.}.\\

\begin{figure*}
    \centering
    \includegraphics[scale=0.8]{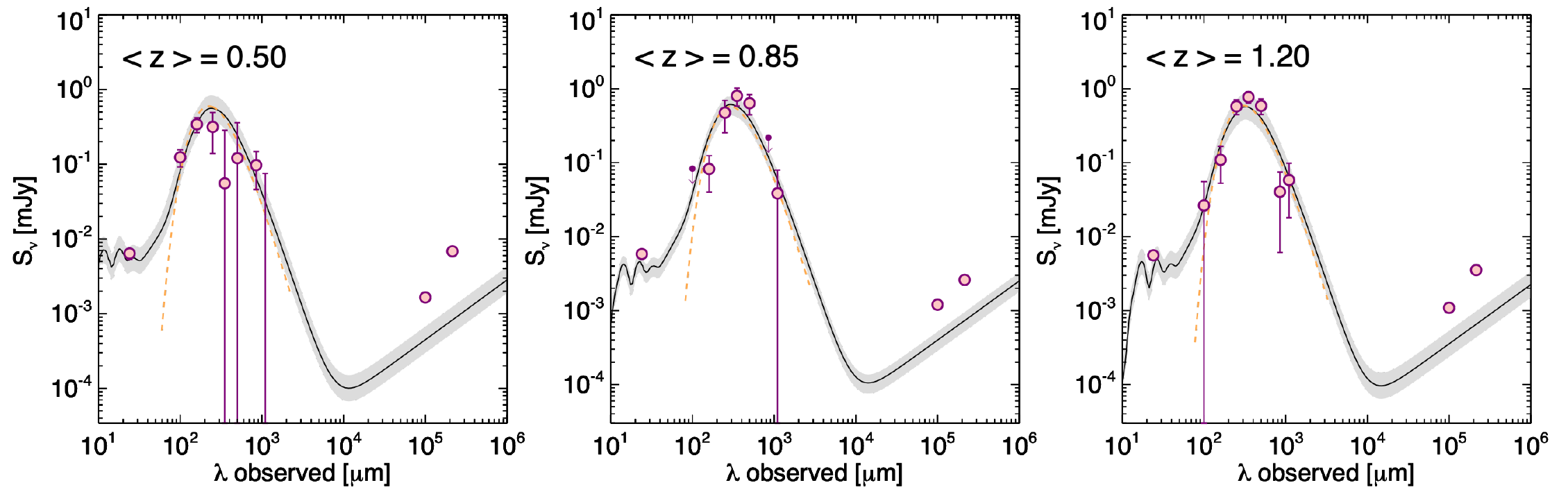}
    \caption{\textbf{Observed median IR SEDs of massive QGs}. The magenta circles mark the observed photometry, with arrows indicating  3$\sigma$ upper limits for bands where the measured flux density was negative. We note that the radio points at 10 and 20\,cm were not included in the modelling. The black line and the grey shaded areas indicate the best-fit model from the \cite{draine_2007} templates and the associated uncertainty. The dashed orange line is the best fit adopting a single \td\ modified black body. The left, middle, and right panels show the results for the redshift intervals $0.3 < z < 0.65$, $0.65 < z < 1.0$, and $1.0 < z < 1.4$, respectively. Salient properties of the sample are reported in Table \ref{tab:properties}.
    }
    \label{fig:seds}
\end{figure*}
\begin{table*}
\small
    \centering
    \caption{Physical properties of the stacked quiescent samples}
    \begin{tabular}{cccccccccc}
    \hline\hline 
        Redshift& $N$& $\mathrm{log}(M_\ast)$& SFR(UV)$\dagger$& $\mathrm{log}(L_{\rm IR})$& \td& $\langle U \rangle$& $\mathrm{log}(M_{\rm dust})$ &$\mathrm{log}(M^{\rm sol}_{\rm gas})$ & $f^{sol}_{\rm gas}$ \\
         & & (\msun)& (\myr)& (\lsun)& (K)& & (\msun)&(\msun)& (\%)\\ 
         (1)& (2)& (3)& (4)& (5)& (6)& (7)& (8)& (9)& (10)\\ 
    \hline
        $0.30<z<0.65$&    $563$&    $11.17\pm0.15$&  $1.7^{+0.3}_{-0.2}$&  $9.40\pm0.05$&  $20.0\pm4.0$&   $0.6\pm0.3$&  $7.51^{+0.20}_{-0.24}$&  $9.46^{+0.20}_{-0.24}$&  $2.0\pm1.0$\\  
        $0.65<z<1.00$&   $1536$&    $11.20\pm0.17$&  $1.7^{+0.3}_{-0.2}$&  $9.89\pm0.05$&  $19.0\pm4.0$&   $0.6\pm0.3$&  $8.01^{+0.20}_{-0.25}$& $9.97^{+0.20}_{-0.25}$&  $5.9\pm3.0$\\  
        $1.00<z<1.40$&    $1394$&    $11.16\pm0.15$&  $1.9^{+0.4}_{-0.3}$&  $10.20\pm0.05$&  $22.0\pm3.0$&   $1.1\pm0.4$&  $8.05^{+0.20}_{-0.24}$&  $10.08^{+0.20}_{-0.24}$& $7.3\pm3.0$\\  
    \hline
    \hline
       $1.50<z<2.20\ddagger$&    $997$&    $11.04\pm0.15$&  $1.5^{+0.4}_{-0.4}$&  $10.35\pm0.13$&  $23.5 \pm 2.0$&   $1.8 \pm 1.2 $&  $8.00^{+0.16}_{-0.19}$&$9.96^{+0.16}_{-0.19}$&  $8.4\pm3.0$\\
    \hline
    \end{tabular}
    \tablefoot{Column 1: Redshift bin. Column 2: Number of objects in the redshift bin. Column 3: mean stellar mass. Column 4: mean SFR(UV). Column 5: total infrared luminosity from the SED fitting with \cite{draine_2007} models integrated over $8-1000\,\mu$m. Column 6: `luminosity-weighted' \td\ from the SED modelling with a single-temperature modified black body. Column 7: mean intensity of the radiation field from \cite{draine_2007} models. Column 8: dust mass. Column 9: molecular gas fraction assuming a solar gas-to-dust conversion factor of \gdr\ = 92. All quantities are computed using a \citet{salpeter} IMF.\\
    $\dagger$: Mean and interquartile range.\\
    $\ddagger$: Stacked photometry from \citet{Gobat_2018} that for consistency was re-fitted here with the same method and setup as for the rest of the redshift bins.}
    \label{tab:properties}
\end{table*}

\begin{table*}
\tiny
    \centering
    
    \caption{Stacked photometry of the QG population, as derived through flux density decomposition.}
    \begin{tabular}{ccccccccccc}
    \hline\hline 
         Redshift& $24\,\mu m$& $100\,\mu m$ & $160\,\mu m$ & $250\,\mu m$ & $350\,\mu m$& $500\,\mu m$ & $850\,\mu m$ & $1100\,\mu m$ & 3.0\,GHz & 1.4\,GHz \\
         &[$\mu Jy$] &[$mJy$]& [$mJy$]& [$mJy$]& [$mJy$]&[$mJy$] & [$mJy$]&[$mJy$]&[$\mu Jy$]&[$\mu Jy$]\\ 
         (1)& (2)& (3)& (4)& (5)& (6)& (7)& (8)& (9)& (10)&(11)\\ 
    \hline
        $\langle 0.50 \rangle$&    $6.40\pm 1.47$&    $0.12\pm0.04$&  $0.34\pm0.08$&  $0.32\pm0.21$&  $0.06\pm0.23$&   $0.12\pm0.23$&  $0.1\pm0.04$&  $<0.225^{\dagger}$& $1.65\pm0.12$& $6.84\pm0.63$\\

        $\langle 0.85 \rangle$&   $5.82\pm 0.61$& $<0.08^{\dagger}$&  $0.08\pm0.04$&  $0.47\pm0.22$&  $0.80\pm0.21$&   $0.64\pm0.19$&  $<0.21^{\dagger}$&  $0.04\pm0.04$&  $1.20\pm0.07$&$2.60\pm0.41$\\   
        
        $\langle 1.20 \rangle$&   $5.60\pm 0.79$& $0.30\pm0.30$&  $0.11\pm0.05$&  $0.57\pm0.13$&  $0.77\pm0.14$&   $0.59\pm0.14$&  $0.04\pm0.03$&  $0.06\pm0.04$& $1.09\pm0.07$ & $3.53\pm0.43$\\   
    \hline

    \hline
    \end{tabular}
        \tablefoot{$\dagger$: $3\sigma$ upper limits following the prescription of \citet{Bethermin_20}}
       \label{tab:photometry}
\end{table*}

On the other hand, the distribution of FIR emission from satellite halos is estimated using multi-wavelength catalogs available for the COSMOS field \citep{muzzin_2013,laigle_2016}. Satellites were selected in the catalogue as objects that both satisfy the $UVJ$ criterion \citep{williams_2009} for SFGs and have compatible photometric redshifts, namely $z_{\text{L}68}<z_{\text{cen}}<z_{\text{H}68}$, where $z_{\text{cen}}$ is the redshift of the central QG and $z_{*68}$ are the lower and upper 68\% confidence limits to the redshift of satellites, respectively. We estimated the SFR of each satellite candidate by fitting its rest-frame ultraviolet SED with constant star formation history stellar population templates \citep{bruzual_2003}, with and without dust extinction. The difference between these two estimates yields an obscured SFR surface density that, when convolved with the instrumental beam in each band, is assumed to be proportional to the FIR flux distribution from the satellite halo. More details are given in Appendix ~\ref{app:satellites}.\\

To extract the FIR SED of the QGs in each redshift bin, we fitted the cutouts in each band with combinations of these two components plus a flat background term. We let the scaling of the satellite emission vary freely at 24\,$\mu$m, 100\,$\mu$m, and 160\,$\mu$m, where the resolution is high enough to properly decompose the observed signal into point-like and extended components.

 Satellite fluxes in these three bands were then fitted with a set of redshift-dependent MS IR templates \citep{magdis_2012} that capture the average value and the scatter of  \td\ and $\langle U \rangle$ of MS galaxies at a given redshift \citep[e.g.][]{magdis_2012,bethermin_2015}. From these templates (and their scatter) we extrapolated satellite fluxes (and their uncertainties) at $\geq$ 250\,$\mu$m, where both the point-like and halo emission are extended and the large beams preclude a non-degenerate decomposition. For these bands, the scaling of the satellite model was fixed to the extrapolated flux, with only the background and central source being fitted. The satellite fluxes thus removed from the FIR SED correspond to $\mathrm{log}(L_{\rm IR}/L_{\odot}) = 10.40\pm0.09, 10.70\pm0.07, {\rm and}~9.30\pm0.10$~for the high-, intermediate-, and low-redshift bins, respectively. These estimates are consistent with the predicted \lir\ infrared from the obscured SFR of the satellites as inferred  by the UV/optical modelling ($\mathrm{log}(L_{\rm IR}/L_{\odot}) = 10.47\pm0.13, 10.77\pm0.09, {\rm and}~9.70\pm0.38$). Finally, we note that while these \lir values are comparable to or higher than the \lir\ of QGs, the sum of the autocorrelation and the QG component still dominates the $\gtrsim$ 250\,$\mu$m bands due to a combination of lower \td\ and large autocorrelation term.

The fits of the various components were done with pixel weights derived from the stacked variance maps described at the beginning of this section. However, these cannot be used to estimate accurate uncertainties on the extracted fluxes as the pixel-to-pixel variance is necessarily correlated in the FIR due to confusion. Therefore, rather than considering the (small) formal errors on scaling of the combined point-source and autocorrelation models, uncertainties on central fluxes were estimated by fitting instrumental beams at random positions in the cutouts, well clear of both the central source and satellite emission. To these we added in quadrature the error of both the satellite fluxes and background, in the corresponding band. We note that the autocorrelation term alone is free of uncertainty by construction, the sample being set and the beam models in $\geq$250\,$\mu$m bands (where it is not negligible) being theoretical and provided without associated errors.
The uncertainties for each decomposition component are presented in Table ~\ref{tab:errors}. 

Finally, the measured central QG fluxes were rescaled down to account for contamination due to random clustering in the image using the corrections described in \citet{bethermin_2015}. We note that the background term might not necessarily be flat, due to lensing from the QGs' host dark matter halo. Magnification bias is typically dominated by area increase, so its net result is a deficit in faint source counts with respect to the field. Assuming that the QGs are central to their halo, this should lead to a lower effective background towards the centre of the cutouts, thus slightly underestimated central fluxes. However, its magnitude depends on the shape of the luminosity distribution of background sources, for which we have little to no information. Furthermore, this effect can be mild even for massive galaxy clusters \citep[e.g.][]{Umetsu_2014}. In this case, where the host halos are smaller by a factor of $\sim$10 (G18), we do not expect it to be significant. The final photometry of our stacked QGs is presented in Table \ref{tab:photometry} while the various decomposition components for each redshift bin and each broadband photometry are presented in Appendix~\ref{app:satellites}.

\section{SED modelling}\label{sec:modeling}
To derive the FIR properties of the stacked samples, we employed the dust models of \citet{draine_2007} (DL07), with the standard  
parameterisation used in previous studies \citep[e.g.][]{magdis_2012}. Namely, we considered diffuse ISM models with radiation 
field  intensities ranging from 0.1 $\leq U_{\rm min} \leq $ 50, while fixing the maximum radiation field to $U_{\rm max} = 10^{6}$.  We allowed for $
\gamma$ (the fraction of dust exposed to starlight with intensities ranging from $U_{\rm min}$ to $U_{\rm max}$) to vary between 
$0 \leq \gamma \leq 0.5$ with a step of 0.01 and  adopted Milky Way dust models with $q_{PAH}$ (i.e. the fraction of the dust mass  in the form of PAH 
grains) from 0.4\% to 4.6\%. To account for the artificial SED broadening due to the redshift distribution of the stacked sources, we constructed an average SED for each DL07 model by co-adding each original DL07 model at each redshift within the bin, weighted by the redshift distribution of stacked sources. The redshift-weighted, `smoothed' models were shifted to the median redshift of each bin and were fit to the data, yielding best-fit \lir, \md, and  \um\ values.

To estimate the uncertainties of the derived parameters we bootstrapped and fitted with same methodology 1000 artificial realisations of the real SEDs, by perturbing the extracted photometry within 1$\sigma$ for the detections (S/N$\geq$3) and within flux $+3\sigma$ for the rest, following the methodology presented in \citet{Bethermin_20}. As uncertainty for each derived parameter from the real data, we adopted the standard deviation of the values of the corresponding parameter inferred  from the perturbed SEDs. As a sanity check we also introduced a bootstrapping component in our analysis by randomly excluding one observed data point at each realisation. We also confirm that the data sets of available detections (which include  at least one data point at $\lambda_{\rm rest} \geq 220\,\mu$m for all four stacked ensembles, as well as a detection at $24\,\mu$m and at least one detection in the Wien part of the SED), along with the upper limits, ensure against  possible systematic effects in the derivation of \md\ and \lir\, for all redshift bins (see Appendix~\ref{app:firsimulation}). Finally, to derive `luminosity-weighted' \td\ we complemented our analysis by considering a single-temperature optically thin  modified blackbody (MBB) models with a fixed effective dust emissivity index of $\beta = 1.8$. 

The best fit models to the data (24\,$\mu$m to 1.1\,mm for DL07 and $\lambda_{\rm rest} > $ 50\,$\mu$m to 1.1\,mm for MBB), along with the range of SEDs from the random 
realisations, are shown in Figure \ref{fig:seds}. The best fit  parameters and their corresponding uncertainties are summarised in Table~\ref{tab:properties}. While the radio data points are not  included in the fit, we add to each the best-fit DL07 model a power-law radio slope with a spectral index $\alpha = 0.8$ and a normalisation given by the FIR–radio correlation 
\citep[e.g.][]{Delhaize17}, to investigate (and visualise) possible radio excess in our stacked ensembles.  

\section{Results}
Using the derived parameters from the SED modelling in the previous section, we attempted to characterise the dust and gas content of QGs and investigate possible evolutionary trends.  For the rest of the analysis we will be referring to the $\langle z \rangle = 0.50, 0.85,$ and $1.20$  stacks as low$-z$, mid$-z$, and high$-z$ quiescent sample respectively. We also complemented our study with the stacked ensemble of G18 at  $\langle z \rangle =1.75$, which has been constructed and analysed using  similar sample selection, stacking, photometry, and SED modelling techniques as those adopted here. Overall, this work covers the average properties of QGs  in the redshift range $0.3 < z < 2.0$ with a homogeneous and systematic approach.
\begin{figure*}
    \centering
    \includegraphics[scale=0.83]{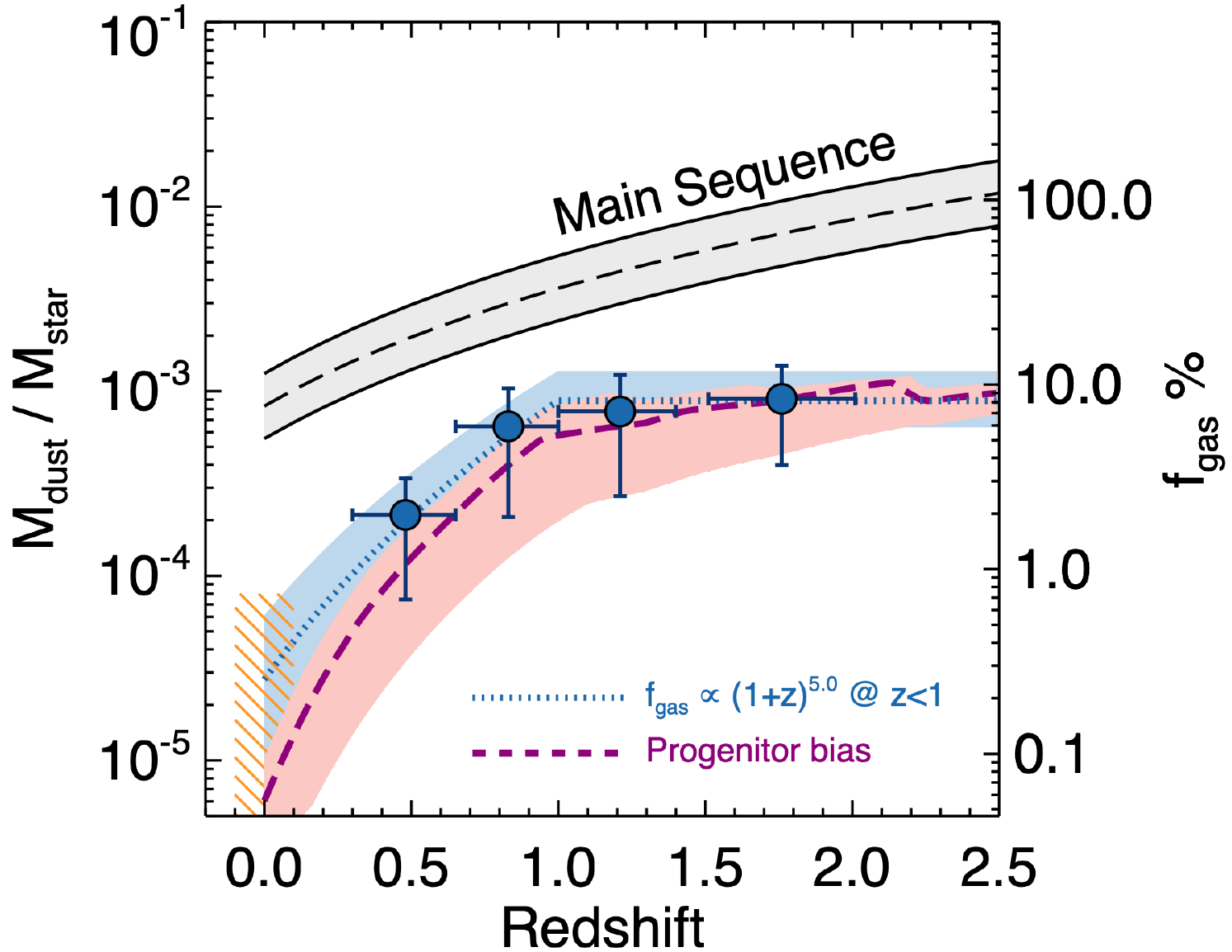}
    \caption{\textbf{Evolution of the dust to stellar mass ratio and of the gas fraction of QGs}. The filled blue circles mark the \md/\mstar\ of the stacked QGs in the $0.3 < z < 2.0$ range, including the results from G18. The grey shaded region depicts the trend for MS galaxies from Kokorev et al.\,(2021) in prep. The orange shaded region captures the range of \md/\mstar\ for local QGs, as drawn from the literature (see text).  The dotted blue line (blue shaded region) corresponds to the best fit  (scatter) to the $0.0 < z < 1.0$ \md/\ms\  data of QGs, with a functional form of \md/\ms\ $\propto (1+z)^{5.0}$. The bold, dashed purple line (shaded region) depicts the evolution (scatter) of \md/\ms\ (and of \fgas) as derived from the progenitor bias analysis of \citet{Gobat_20}, which is discussed in Section 5.2 and Section 6. The conversion of \mdust/\mstar\ to \fgas\ assumes solar metallicity and \gdr\ = 92.}
    \label{fig:mdms}
\end{figure*}
\subsection{Basic FIR properties}
The available mid-IR to millimetre photometry that spans from observed 24- to 1100\,$\mu$m allows for a robust determination of the total 8-1000\,$\mu$m infrared 
luminosities (\lir), dust masses (\md), mean radiation field 
(equivalent of dust mass-weighted luminosity or temperature), parametrised as  $\langle U 
\rangle \propto$ \md/\lir, and  luminosity-weighted dust 
temperatures (\td). Our SED fitting yields low infrared 
outputs for our stacked ensembles at all redshifts with 
$\mathrm{log}(L_{\rm IR}/L_{\odot}) = 9.40 - 10.30$, dust 
masses in the range of 
$\mathrm{log}(M_{\rm dust}/M_{\odot}) = 7.50 - 8.15$, low mean
radiation field  $\langle U \rangle = 0.6-1.2$, and cold \td\ $< 23$\,K dust temperatures. 

In comparison to previous studies that presented stacked SEDs of QGs at similar redshifts and equivalent \ms\ bins, our analysis yields significantly lower \lir\ and \md\ estimates. For example, fitting the photometry presented in \cite{Man16} with same methodology as for our data, we find $\mathrm{log}(L_{\rm IR}/L_{\odot}) = 9.8-11.2$ and $\mathrm{log}(M_{\rm dust}/M_{\odot}) = 8.0-8.6$ for the stacked ensembles of QGs at $z = 0.4 - 1.6$. These values are $0.2-0.9$\,dex larger than the corresponding parameters inferred from our analysis for our sample. 
These discrepancies stem from the more conservative sample selection that we adopted in this work (double-colour criterion and  exclusion of individually detected 24\,$\mu$m sources). In addition, the adopted stacking and, especially, the treatment and correction for the contribution of blended satellites further affects the comparison with \citet[see also G18]{Man16}. Nevertheless, and in agreement with \citet{Man16}, we do find a radio excess between the observed radio fluxes and those inferred by the infrared-radio correlation of SFGs by \citet{Delhaize17}, indicative of AGN activity and  `radio-mode' feedback in massive QGs \citep[e.g][]{Gobat_2018,barisc17}.  

In order to test for the possible sensitivity of these recovered parameters to the selection criteria described in Section~\ref{sec:selection}, as well as possible contamination of quiescent samples by dusty star forming objects, we performed a series of tests using sub-samples obtained by either applying an additional 0.1\,mag colour margin or by dividing the low-, mid-, and high-$z$ bins into blue and red halves. Applying the same stacking, source extraction, and modelling procedures, as described in Sections~\ref{sec:method} and \ref{sec:modeling}, to these yields SEDs that are entirely consistent with the full low-, mid-, and high-$z$ samples. This suggests that our analysis is robust with respect to selection effects. More details are given in Appendix~\ref{app:subsamples}.

Finally, we inferred gas masses (\mgas) estimates by converting the derived \md\ to \mgas\ through the metallicity-dependent dust to gas mass ratio () equation of \citet{magdis_2012}; $\rm log(GDR) = 10.54 - 0.99\times(12+log(O/H))$. We considered a range of metallicities $Z_{\odot} \leq Z \leq 2Z_{\odot}$ (8.66 $\leq$ 12$+$log(O/H) $\leq$ 9.1 in the \citealt{Pettini04} scale), appropriate for the SFR and \ms\ of our samples \citep[e.g.][]{Mannucci10}. For these metallicities the GDR varies between $35 \leq$ \mgas/\md\ $\leq 92$. For the rest of our analysis we use the \mgas\ estimates based on a solar metallicity as benchmark and those for a super-solar metallicity as lower limits. Therefore, throughout the paper, \mgas\ = $M^{\rm sol}_{\rm gas}$ (and \fgas\ = $f^{\rm sol}_{\rm gas}$), unless otherwise stated. We note that since the \md/\mgas -- $Z$ relation of \citet{magdis_2012} is almost linear (slope of 0.99), we can easily rescale \mgas\ estimates between solar  and an arbitrary metallicity Y by using:
\begin{equation}
	\mathrm{log}[M^{Y}_{\rm gas}] = \mathrm{log}[M^{\rm sol}_{\rm gas}] + [8.66 - Y],\,\,\,\, Y = 12 + \mathrm{log(O/H)} . 
\end{equation}

We also recall that the GDR -- Z relation yields \mgas\ estimates that trace the sum of the atomic ($M_{\rm HI}$) and the molecular ($M_{\rm 
H_2}$) gas mass (including a $\times$1.36 He contribution). As such, \mgas\ = $M_{\rm HI}$ + $M_{\rm H_{2}}$. 
While the exact state of the cold gas in our QGs cannot be 
characterised, the $\langle M_{\rm H2}/M_{\rm HI} \rangle > 1$ ratios
 reported in the literature for $z > 0.5$ and for high stellar surface densities \citep[e.g.][]{Blitz06, Bigiel08, Obreschkow09} motivate us to make the simplified assumption that $M_{\rm H_{2}} >> M_{\rm HI}$ and, thus,   $M_{\rm H_{2}} \approx$ \mgas. We note though, 
that this estimate is an upper limit to the molecular gas mass of our samples.

\begin{figure*}[!t]
    \centering
    \includegraphics[scale=1.1]{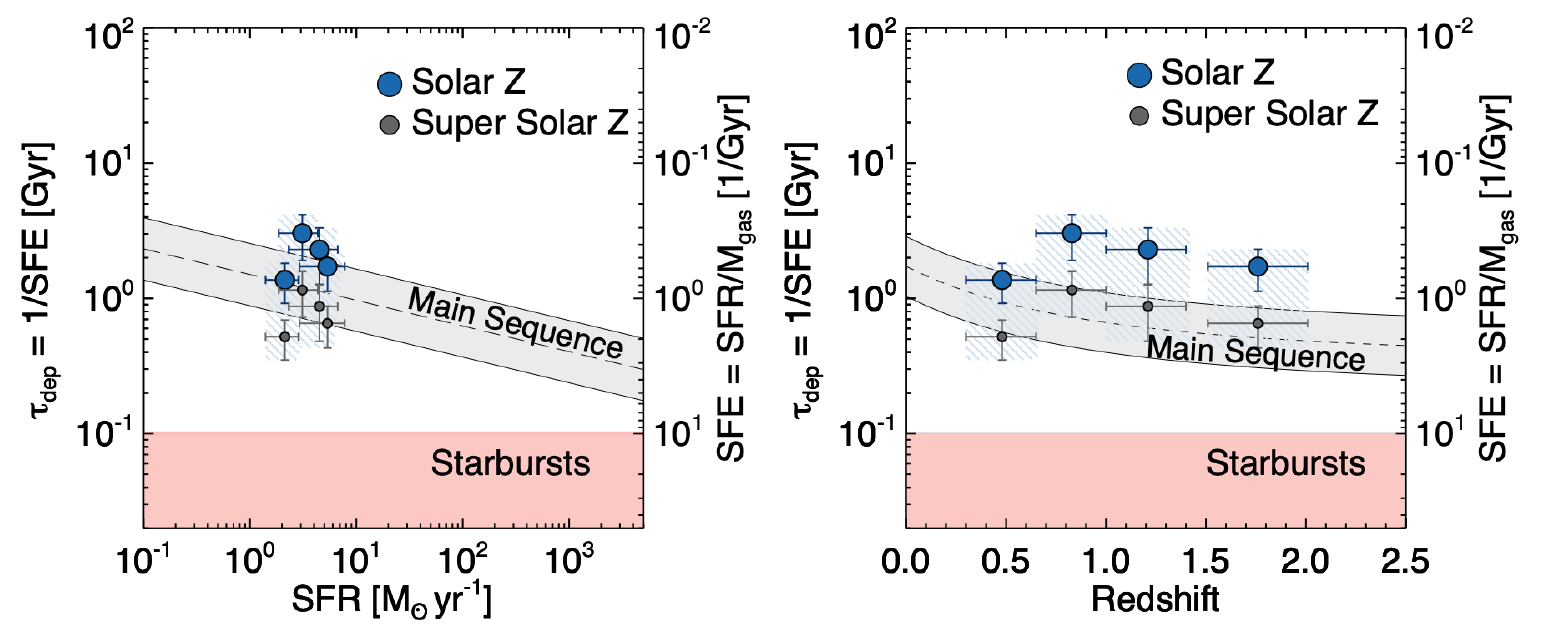}
 \caption{\textbf{Gas depletion timescales}. \textit{Left}: Gas depletion time scale (\tdep) as a function of SFR. The blue (grey) circles correspond to the stacked samples of QGs for the case of solar (super 
 solar) metallicity, assuming GDR = 92 (35). The dashed grey line and the grey shaded region depict the trend for MS galaxies from \citet{Sargent14} with a scatter of $0.23$\,dex. The magenta shaded region marks the area with \depl\ $<$ 100Myrs, typical of star-bursting galaxies. \textit{Right}: Gas depletion time scale as a function of redshift. The symbols are the same as in the left panel. The dashed grey line and the grey shaded region depict the trend of \citet{Liu19} for MS galaxies at fixed \mstar\ = $2\times10^{11}$\msol\ and its corresponding scatter of $0.23$\,dex.}
\label{fig:tdep}
\end{figure*}

\subsection{Evolution of dust and gas mass fractions}
In Figure \ref{fig:mdms} we bring together the \md\ estimates for our stacked ensembles as well as a range of literature \md\ estimates for local QGs of comparable \ms\ \citep{Smith12,  Boselli14, Lianou16, Michal19} and explore the evolution of dust mass fraction, \md/\ms, as a function of redshift. Our data indicate that the \md/\ms\ remains roughly constant from $z=2.0$ to $z=1.0$, followed by sharp decline towards the present day, with a functional form of \md/\ms\ $\propto (1+z)^{5.00^{+1.94}_{-1.55}}$ (for $z < 1.0$).  As shown in Figure \ref{fig:mdms}, a similar trend characterises the evolution of the dust mass fraction in SFGs (Kokorev et al.\,2020 in prep.), albeit with a shallower slope. Evidently, QGs exhibit systematically lower dust mass fractions with respect to SFGs at all redshifts, indicative of dust consumption, destruction or expulsion during the quenching phase.  However, the flat evolution for QGs between $z = 2.0 - 1.0$ and the monotonically decreasing \md/\ms\ for SFGs at all redshift, results in a minimum deviation between the two populations at $z \sim 1.0$. 

The evolution of \md/\ms\ we find here also mirrors the evolution of the gas fraction of QGs. Under the assumption of constant solar metallicity for massive QGs at all redshifts, we can directly convert the  \md/\ms\ presented in the left y-axis of Figure \ref{fig:mdms} to  \fgas. This approach yields a constant \fgas\ $\approx 7-8\%$ between $z = 2.0 - 1.0$ that subsequently drops to \fgas\ $\approx 2\%$ at $z = 0.5$ and  \fgas\ $\leq 1\%$ in the local Universe. These values are in agreement with independent \mgas\ estimates or upper limits inferred by CO observations of individual QGs and post-starburst galaxies at various redshifts \citep[e.g.][]{French15, Sargent15, Rudnick17, Suess17, Spilker18,  Bezanson19}.  We note that for a super solar metallicity of $2\times Z_{\odot}$ the \fgas\ values presented in Figure \ref{fig:mdms} would correspond to $f^{\rm Z}_{\rm gas} \approx 0.4 \times f^{\rm sol}_{\rm gas}$. In Section 6, we will interpret the observed redshift evolution of \md/\ms\ and \fgas\ of QGs in terms of progenitor bias and ageing of the stellar population.

\subsection{SFR and gas depletion time scales }
The gas masses inferred in the previous section, when coupled with the corresponding SFRs of the stacked ensembles, can be used to define the 
star formation efficiency (SFE = SFR/\mgas) or, inversely, the gas depletion time scale (\depl\ 
$=$ 1/SFE).  To derive the total SFR in each redshift bin,  we 
considered the average unobscured component as derived from the optical photometry (SFR$_{\rm 
UV}$) and the obscured star formation  by converting \lir\ to SFR$_{\rm  IR}
$ using the \citet{Kennicutt98} scaling relation. We caution the reader that for QGs a 
considerable fraction of \lir\ could arise from dust heated by old rather than 
young stellar populations \citep[e.g.][]{Hayward_14}, as assumed by the Kennicutt SFR-\lir\ relation.  Therefore, 
the estimates of SFR$_{\rm tot}$ = SFR$_{\rm IR}$ + SFR$_{\rm UV}$ should be regarded as upper 
limits. To constrain the fraction of \lir\ arising from dust heated by old stars, we fit 
each galaxy's optical-NIR SED with composite stellar population models derived from 
\citet{bruzual_2003} templates. For simplicity, we used delayed exponential star formation 
histories, starting at $z_{\rm in}=10$ and with time scales $\tau=100\,\rm Myr-3\,\rm Gyr$, each 
truncated 1\,Gyr before observation. Finally, we assumed a \citet{noll_2009} attenuation curve 
with variable slope and computed the luminosity difference of the best-fit model before and 
after attenuation. We find that this absorbed energy does not make up more than $\sim$30\% 
of \lir\ at $z\gtrsim1$ but could represent up to 100\% of the infrared luminosity of $z\sim0.5$ 
QGs. Given the uncertainties and the small IR output indicated by our stacks, we chose to consistently consider SFR$_{\rm tot}$ at all redshift bins.

The emerging (maximum) SFRs are listed in Table~\ref{tab:properties} and they range from $\sim$2.0\,\sfrunit\ in the lowest redshift bin ($\langle z \rangle  = 0.5$) to $\sim$6.5\,\sfrunit\ at $\langle z \rangle = 1.8$.  These average SFRs place our stacked ensembles $\geq$10$\times$ below the MS at their corresponding 
redshifts, confirming their passive nature. At the same time, \depl, ranges from 1.3 to 3.0\,Gyr for the case of solar metallicity with a mean $\langle$\depl$\rangle$ = $2.1\pm0.8$\,Gyr  and 0.5 to 1.1\,Gyr for a super solar metallicity.  These \depl\ estimates are plotted as a function of SFR and redshift in the two panels of Figure \ref{fig:tdep} along with the corresponding tracks for MS galaxies \citep{Sargent14,Liu19}. We stress again that since we are using the maximum SFR estimates the quoted \depl\ should be regarded as lower limits for each metallicity. In agreement with \citet{Spilker18}, we find that distant QGs, for their SFRs, have a \depl\ that is broadly consistent with that of local MS galaxies, but longer than that of MS galaxies at their corresponding redshift. We note that  while our data hint upon an increase in \depl\ from $z=2.0$ to $z=0.8$ followed by a decline towards later cosmic epochs (Figure \ref{fig:tdep}, right), the values are overlapping within their uncertainties yielding a non significant correlation. Possible metallicity effects as well as the uncertainties linked to the conversion of \lir\ to SFR, will be investigated in detail in a future study.

\subsection{SED and \td\ evolution }
The star formation activity and the dust heating photons per unit ISM mass, both parametrised above in terms of SFR, \md, and SFE,  should also be imprinted in the shape of the FIR SED and the characteristic \td\ of QGs. Indeed, a large body of literature that has focused on the evolution of \td\ as a function of redshift for MS and star-bursting galaxies alike \citep[e.g.][]{Hwang10, magdis_2012,magdis17, Magnelli14,schreiber_2018_dust, Jin19, Cortzen20} has established that the increase in sSFR and SFE at higher redshifts is also followed by a trend of warmer SEDs and increasing \td\ at least out to $z\sim4$. At the same time, SEDs and the effective \td\ of distant QGs as a function of redshift are far less explored.  

\begin{figure*}
    \centering
    \includegraphics[scale=1.1]{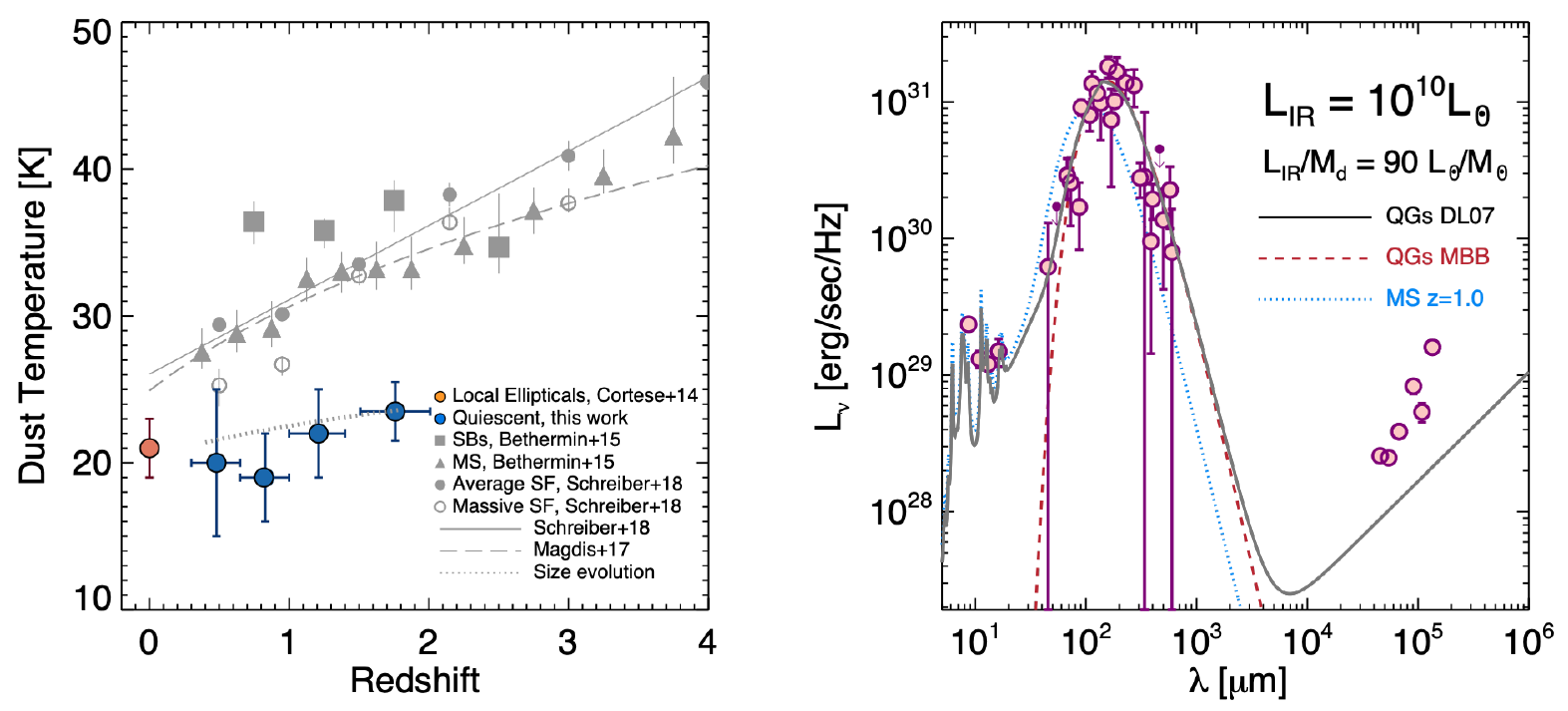}
    \caption{\textbf{Dust temperatures and average SED}. \textit{Left}: The evolution of \td\ in 
    QGs. The filled blue circles mark the luminosity-weighted \td\ for the stacked QGs at $0.3 < z < 2.0$, including the results from G18. Filled grey triangles and squares indicate SFGs on the MS and strong starbursts from \cite{bethermin_2015}.  Filled and open grey circles represent the mass-averaged values and the most massive galaxies ($11.0 < \mathrm{log}(M_\ast/M_\odot) < 11.5$, Salpeter IMF) for the star-forming sample of \cite{schreiber_2018_dust}, respectively. The solid and dashed grey lines indicate the best fit models from \cite{schreiber_2018_dust} and \cite{bethermin_2015}, extending the parameterisation of \cite{magdis_2012}. The dotted grey line indicates the expected \td\ track for a size evolution of QGs,  $R \propto (1+z)^{-1.48}$. The track is normalised to the \td\ of the $z = 1.8$ sample. \textit{Right}: Composite and template FIR SED for $z > 0$ QGs. The magenta circles (arrows) correspond to the photometry (3$\sigma$ upper limits) of the stacked ensembles presented in Figure~\ref{fig:seds}  normalised to \lir\ = 10$^{10}$\,\lsol\ (and \md $= 1.11\times10^{8}$\,\msol).  The DL07 and MBB model fits to the SED are shown with grey and red dashed lines. The dotted blue line correspond to the typical SED of $z = 1.0$ MS galaxies from \citet{magdis_2012}, normalised to the same \lir. The QG template is publicly available at \href{www.georgiosmagdis.com/software/}{www.georgiosmagdis.com/software/}.} 
    \label{fig:temp}
\end{figure*}

In Figure \ref{fig:temp} we plot the \td\ of  QGs derived from our MBB fits as a function of redshift, along with literature data and trends for MS galaxies. As expected, QGs exhibit colder \td\ with respect to star forming galaxies at all redshifts, indicative of a lower star formation activity.  Also, at first reading, the \td\ estimates for our quiescent ensembles are overlapping  within their uncertainties, with an average $\langle T_{\rm d} \rangle = 21.0 \pm 2.0$\,K.  Notably, in the redshift range probed by our data, SFGs have witnessed a decrease in their \td\ by $\sim8$\,K, indicative of a  steep evolution that is not observed for QGs. 

However, the hint of small \td\ decrease in QGs between $z = 2.0$ and $z = 0.3$ ($\Delta$\td\ $\approx 
3$\,K), even if not statistically vigorous,  provides the grounds for a tempting speculation of a weak 
evolution driven by their size evolution. Indeed, a size evolution of $R \propto (1+z)^{-1.48}$ 
\citep{vdw14} yields a size increase of a factor of $\sim0.5$ between $z = 1.8$ and 0.8. Similarly, 
following the analysis of  \citet{Chanial07}, $T^{4+\beta}_{\rm d}$ $\propto$ $[\Sigma_{\rm 
IR}]^{\gamma}$, where $\gamma=\frac{0.4}{1.4}$ and $\Sigma_{\rm IR} \propto L_{\rm IR}/R^{\rm 2}$. 
For fixed \lir\ (or SFR), as in the case for our stacked samples in this redshift range, it follows that 
$T_{\rm d}(z_1)$/$T_{\rm d}(z_2)$ $
\propto [R(z_2)/R(z_1)]^{2 \times \gamma/(4+\beta)}$. Anchoring this equation at $T_{\rm d}(z=1.8) = 23.5$\,K, as inferred by our analysis and adopting $\beta=1.8,$ provides an evolutionary track that is in broad agreement with our measurements for the $0.3 < z < 2.0$ stacks. We conclude that a weak (if any) evolution of \td\ in QGs can be understood in terms of size growth. At the same time an average $\langle T_{\rm d} \rangle = 21.0 \pm 2.0$\,K provides a statistically robust estimate for a characteristic \td\ for the ISM of QGs over the last ten billion years. 

In addition, the small variance in \td\ and \lir/\md, suggests  that  massive QGs at various redshifts  share a common FIR SED shape, albeit with a varying normalisation that mirrors the evolution of  \lir\ and of the ISM mass budget as described in the previous sections. To produce an average SED, we bring the stacked photometry of each redshift bin at rest-frame, normalise it so that all SEDs are anchored at the same reference luminosity of $1.0 \times 10^{10}$\,\lsol\ (or SFR $\approx 1.7$\,\sfrunit), and fit with DL07 and MBB models following the methodology presented in Section 4. The observed points (and upper limits) along with the best-fit models are presented Figure \ref{fig:temp}\,(right). The emerging average SED of QGs has  \td\  $= 20$\,K, and \lir/\md\ $\approx 90\,$\lsol/\msol, in direct contrast to that of typical $z > 0$ star forming galaxies. As an example, we consider the average SED of $z = 1.0$ MS galaxies from \citet{magdis_2012}, which clearly peaks at shorter $\lambda$ with a \td\ $\approx$ 30\,K and has approximately a factor of $\sim10\times$ larger infrared energy output per unit dust mass (\lir/\md\ $\approx$ 1000\lsol/\msol). Finally, we stress that the radio data are not included in the fit but are presented to demonstrate the  radio-excess that has already briefly discussed in a previous section. The average SED of QGs (normalised to \lir\ = $1.0 \times 10^{10}$\,\lsol\, and \md\ = 1.11 $\times$ 10$^{8}$\,\msol) is publicly available{\footnote{http://www.georgiosmagdis.com/software/}}.

\section{Discussion}
In the previous sections, we characterised the mass budget and the  properties of the ISM of QGs at various redshifts. Here we  offer an explanation for the origin and the drivers of the observed trends.

To interpret the evolution of \md\ in our quiescent sample, we first need to consider the expected redshift evolution of a population of QGs. At log(\ms/\msoll) $\sim 11.0$, their co-moving number density increases significantly between $z \sim 4$ and $z \sim 1$, with little evolution after that \citep[e.g.][]{Muzzin13,Davidzon17}. This implies that an unbiased sample of QGs will contain a large fraction of newly formed objects at any redshift $z \gtrsim 1$ while, few new objects enter the $z < 1$ population. For simplicity, let us neglect dry merging events and equate the formation time of a QG with the quenching time of its precursor. In this case, the average age of the quenching event will vary little between quiescent galaxy samples at $z\gtrsim1$ and increase with cosmic time at $z\lesssim1$. In this context, the apparent non-evolution of the dust fraction between $z\sim2$ and $z \sim1 $, where newly quenched galaxies constitute a large part of the quiescent population, suggests that whatever quenching mechanisms are at play, they leave behind very similar ISM conditions independently of redshift. Conversely, the steep decrease in the dust fraction between $z \sim 1$ and $z = 0$ is consistent with a consumption (or destruction) without replenishment of the dust and gas as few to no new QGs enter the massive population in this redshift range. Combining both regimes yields an apparent evolution very similar to our observed constraints, as shown in Figure~\ref{fig:mdms}. Here we estimated the median gas fraction of the whole population of QGs as a function of redshift by assuming only the QG production rate from stellar mass functions, an initial gas fraction (constant with redshift) of $\sim$10\% immediately following quenching, and a closed-box consumption with a time scale of $\sim$2\,Gyr after that \citep{Gobat_20}. The track derived in this manner reproduces our data well, especially when considering that the latter are likely slightly biased towards dustier objects at lower redshift (see Figure~\ref{fig:24um_detections}), which would flatten the observed trend.

This `progenitor-bias' explanation for the observed evolution of \fgas\ and \md/\ms\ would also require an anti-correlation between these quantities and the average quenching time, or more simply, the stellar age.  Indeed, in a recent study, \citet{Michal19}, presented an exponential decline of the \md/\ms\ with age for a sample of galaxies at a fixed cosmic epoch ($\langle z \rangle = 0.13$) and a range of luminosity-weighted stellar ages between $\rm 9.0 < log(age/yr) < 10$. The authors report a decline of \md/\ms\ with age that follows $e^{\rm -age/\tau}$ and a dust lifetime $\tau \approx$ 2.5\,Gyr.  While we do not have any information about the actual stellar age of the galaxies in our stacks, the effect of varying age can be mimicked by the our different redshift bins and under the assumption that the passive galaxies at $z > 0.9$ will, on average, remain passive at later times. Indeed the time interval between $z = 0.9$ and $z = 0.5$ corresponds to $\approx$2.3\,Gyr. According to the formula of \citet{Michal19}, and assuming an average  stellar age of $2 - 3$\,Gyr for the QGs at $z = 0.9$, we should expect a drop in  log(\md/\ms) by $\sim0.4-0.6$\,dex, in very good agreement with our observations. At the same time, as described above, the flat behaviour of the \md/\ms\ (and \fgas) at $z > 1.0$ could then be explained by the younger, and roughly constant stellar age of the QGs in the higher redshift bins (e.g. log[Age/yr] = 9.0 at $z = 2.0 - 4.0$, \citet{toft17,V20}). 

Finally, the rather constant FIR SED shape (and \td) of QGs at all redshifts, indicate that \lir\ and \md\ roughly scale together  irrespective of the time of quenching. If indeed \lir\ is a proxy of SFR and \md\ a proxy of \mgas\ this means that the decrease in \mgas\ is closely followed by a decrease in SFR. In this case, the universality of the initial ISM conditions, as advocated by the evolution of ISM mass budget (in terms of \fgas\ and \md/\ms),  is followed by a regularity in the internal processes within quenched galaxies.

We should stress that while the progenitor bias offers an appealing empirical explanation for the observed evolution of the ISM properties of QGs, it bares little, if any, information about the underlying physics of the origin and the maintenance of 
quenching. For the  latter, the radio excess present in our SEDs as well as in previous studies \citep[e.g.][]{Man16, DEugenio20}, 
point towards `radio-mode' feedback from AGN-driven jets as compelling quenching mechanism that can both prevent gas accretion due to 
heating of the halo gas and expel a fraction of the cold gas reservoir \citep[e.g.,][]{croton_2006, Gobat_2018}. This radio-mode 
feedback would also need to be stronger at higher redshift to compensate for the higher (putative) gas fractions seen at these epochs. Using 
the average 1.4\,GHz excess, with respect to the FIR-radio correlation, and the evolution of the AGN luminosity function \citep{Novak_2018}, 
we indeed estimate that the AGN duty cycle for our QGs decreases from $>$50\% at $z\sim1.8$ (see G18) to $\sim$15\% in the 
low-$z$ redshift bin (see Appendix~\ref{app:dutycyle}).

\subsection{Outlook and ALMA predictions}
From the discussion above, it stems that the \md\ (and subsequently \mgas\ and  \fgas) of QGs depends both on redshift and on time elapsed since the quenching event. So while our \md\ and \mgas\  estimates should be valid for the average QGs population at a given redshift, this might not be the case for individual sources due to possible age variations. Clearly, the way forwards necessitates focused ISM studies for large and representative samples of individual QGs at various cosmic epochs. In this regard, the IR SED template presented in Section 5, can facilitate a series of useful predictions regarding the observability of their dust continuum emission.

The main characteristic of the template is the ratio of its native infrared luminosity ($L^{\rm 0}_{\rm IR}$) to its native dust mass ($M^{\rm 0}_{\rm dust}$), which is expressed as $L^{\rm 0}_{\rm IR}$/
$M^{\rm 0}_{\rm dust}= 90$ \lsol/\msol. Consequently, for any arbitrary normalisation of the template, it follows that for this template, \lir\ = 90$
\times$ \md\ [in \lsol]. Similarly, \mgas, for a fixed metallicity $Z$, scales linearly with \md, with \mgas\ = GDR($Z$) $\times$ \md\ \citep[e.g.][]{Leroy11, magdis_2012}. Therefore, the template can be rescaled to any arbitrary \md\ that corresponds to a given \mgas\ (or \fgas). Indeed,
since \fgas\ = \mgas/\ms\ =  GDR(Z) $\times$ \md/\ms\, we simply have to 
multiply the SED by $N=$ \md/$M^{\rm 0}_{\rm dust}$, which takes the form of:
\begin{equation}
N = \frac{f_{gas} \times M_{\ast}}{GDR(Z)\times M^{\rm 0}_{\rm dust}} .
\end{equation}

Assuming that our SED is representative of massive QGs with a typical log(\ms/\msoll) = 11.2 at all redshifts and by adopting a universal solar metallicity that corresponds to GDR(Z$_{\odot}$) = 92 \citep[e.g.][]{Leroy11, magdis_2012}, the only free parameter in the SED normalisation is \fgas.  Then we  can use our normalised template SED to calculate the emitted flux density  at the central wavelength of various ALMA bands at various redshifts for any adopted \fgas\, after  implementing CMB effects as prescribed in \citet{daCunha13}.
\begin{figure*}[!t]
    \centering
    \includegraphics[scale=0.85]{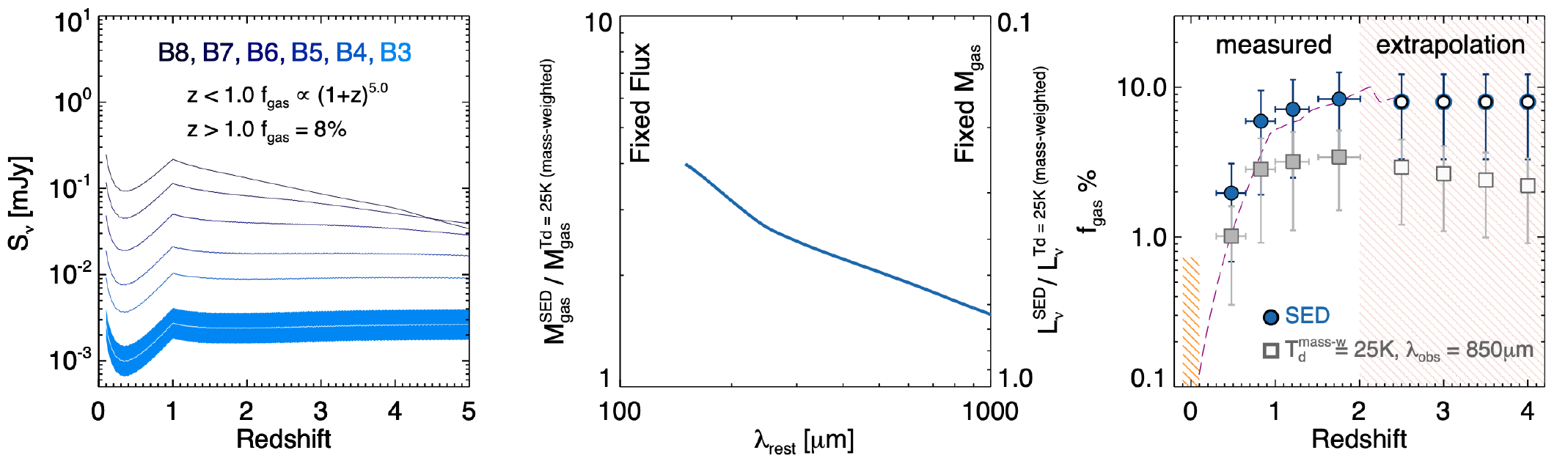}
    \caption{\textbf{ALMA outlook and systematics}. \textit{Left:} ALMA flux densities in various bands for an evolving  \fgas\  as $(1+z)^{5.0}$ at $0.0 < z < 1.0$ and fixed  \fgas\ = 8\%, at $z > 1.0$. The predicted fluxes are derived using the average DL07 SED presented in Figure \ref{fig:temp}(right) and are  corrected for CMB effects.  The blue shaded region depicts the range of predicted flux densities in ALMA Band 3 assuming a 50\% uncertainty in the adopted \fgas.  \textit{Middle:} Ratio of \mgas\ estimates, as inferred for a fixed 
    monochromatic luminosity at a given rest-frame wavelength  from 1)  scaling the 
    average SED of QGs presented in Figure \ref{fig:temp}(right), 
    $M^{\rm SED}_{\rm gas}$, and 2) the prescription of \citet{Scoville_2017}, $M^{\rm Td=25K\, (mass-weighted)}_{\rm gas}$. Similarly, the right y-axis 
    depicts the ratio of the monochromatic luminosities 
    predicted by the two methods for a fixed \mgas. \textit{Right:} The effect of the adopted  \mgas-method  on the recovered evolution of \fgas. Filled blue circles correspond to the values presented in Figure 4 using the full SED analysis ($f^{\rm SED}_{\rm gas}$).  Filled grey squares correspond to the values derived from the \citet{Scoville_2017} prescription, assuming monochromatic observations at $\lambda_{\rm obs} = 850 \mu$m, ($f^{\rm 850}_{\rm gas}$). The empty points (and the  colour-shaded region) are an extrapolation to higher redshifts under the assumption of fixed $f^{\rm SED}_{\rm gas}$ = 8\% at $z > 2.0$.}
    \label{fig:alma}
\end{figure*}

As an example, we examine the case where \fgas\ is evolving with redshift as \fgas\ $\propto(1+z)^{5.0}$ at $z < 1.0$ and then remains flat with \fgas\ = 8\% at $z \geq 1.0$ (Section 5).  The flux density tracks  as a function of redshift for various ALMA bands are presented in Figure \ref{fig:alma}(left). We stress that these tracks are specifically tailored around massive (log(\ms/\msoll) = 11.20) QGs with a solar metallicity. However, the recipe described above is also valid for any arbitrary gas phase metallicity and any \ms\ provided that our template is also representative  of less (or more) massive QGs.

We emphasise that the conversion of R-J flux densities ($f_{\rm ALMA}$) to \mgas, and vice versa, is heavily dependent on the adopted method (and template). Indeed, at fixed $f_{\rm ALMA}$, the commonly adopted monochromatic prescription of \citet{Scoville_2017} yields systematically lower \mgas\ estimates with respect to those inferred in here. As shown in  Figure \ref{fig:alma}(middle) the tension between the two methods becomes more prominent at $\lambda_{\rm rest}$ closer to the peak of the SED and progressively eases moving further into the R-J tail. This is also demonstrated in Figure \ref{fig:alma}(right) where we show how applying the \citet{Scoville_2017} method at $\lambda_{\rm obs}= 850 \mu$m would affect the recovered evolution of \fgas. We note that while we lack observations at $z > 2.0$ we extrapolate to higher redshifts under the assumption of flat \fgas\ evolution at $z > 2$ to highlight how discrepancy (for fixed $\lambda_{\rm obs}$)  between the two methods increases at higher redshifts.

The tension in the \mgas\ (and \fgas) estimates between the two methods, arises predominantly from the assumed \td\ of QGs. We recall that the \citet{Scoville_2017} recipe is calibrated on SFGs, assuming a fixed `mass-weighted' \td\ $=$ 25\,K. This corresponds to a $\sim$7-10\,K warmer `luminosity-weighted' \td\ compared to that of our template. Similarly, as discussed above, our QG template is $\sim$10\,K colder with a $\sim$10 times lower \lir/\md\ compared to that of a typical SFG at $z = 1.0$.
It then naturally follows that with respect to the \citet{Scoville_2017} approach (or a $z >$ 0 SFG template) our method predicts lower $f_{\rm ALMA}$ and 
subsequently a significantly  longer integration time to detect dust emission from a QG of a given \mgas\ (e.g. $\sim \times6$ for a  galaxy at $z = 3$ and $
\lambda_{\rm obs}$ = 1200$\mu$m). This, could explain the scarce dust continuum detections among the limited number of high$-z$ QGs that have 
been followed up with ALMA so far.

Another point of concern is selection biases, especially in terms of  \mstar. \citet{Williams_20}, reports no CO[2-1] detection for a sample of 
massive $\langle log(M_{\ast}/M_{\odot}) \rangle  \approx 11.7$ (for Salpeter IMF) QGs at $z \sim 1.5$, placing stringent upper limits in their \fgas $\leq$ 1.0-3.5\%, that are lower compared to our estimates.  However, we note that this sample is $\sim$0.5-0.7dex more massive with respect to the stacked ensembles studied here. If the \fgas\ of high-$z$ QGs decreases as a function of \mstar, similarly to what is observed for SFGs and local ETGs \citep[\fgas\ $\propto M^{-(0.40-0.55)}_{\ast}$ e.g.][]{Saintonge11a1, magdis_2012}, then rescaling our \fgas\ estimates accordingly would significantly ease the tension \citep[see also Fig. 6 of][]{Williams_20}. Nevertheless, we caution the reader that 1) the dust method to infer \mgas, when applied to QGs, could in principle recover a sizeable amount of 
HI that is untraceable by the CO lines and 2) DL07 or similar dust emission models  could be a poor representation of the dust grain composition of distant 
QGs.

Finally, we discuss the impact of the CMB on the predicted flux densities, an effect that was recently observationally confirmed in a sample of `cold' $z > 3$ SFGs \citep{Jin19,Cortzen20}. Our analysis suggests that while $f_{\rm ALMA}$ should remain unaffected up to $z \sim 2.0$ in all ALMA bands, the `dimming' of the dust continuum emission that can be measured against the CMB is not negligable at higher redshifts and lower frequencies. This is demonstrated by the flat evolution of the flux density tracks in Bands 3 $-$ 6,  that clearly deviate from the expected rising trend due to the negative $k-$correction (Figure \ref{fig:alma}left).  Indeed, our analysis suggests that the recoverable continuum emission  drops by a factor of 1.4 at $z$ = 3 - 5 for Band 5 and by a factor of  $\sim$1.6-2.0  at $z =$ 4-5 for Band 3 observations. Admittedly, the analysis presented in this work does not extent beyond $z \sim 2$, and therefore the actual IR SED (and \td) of QGs at $z > 2$ is still unconstrained.  However, if the template SED presented here is also valid at $z > 2.0$, CMB-corrections should be implemented both in the observing strategy and in the interpretation of dust continuum and spectral line observations of $z > 3.0$ QGs in the (sub)-mm.

\subsection{Caveats} 
As pointed out in the previous sections, this analysis is not free of caveats which are inherent to the stacking method. First, decomposing the stacked signal into various components is a challenging task. Indeed, `satellite' fluxes in the SPIRE bands, and their corresponding uncertainties, depend on the adopted templates used for extrapolation. While we have considered an uncertainty on the SED of the SF component, our analysis does not account for spatially correlated variations of \tdust. This would effectively make the satellite emission term frequency-dependent. On the other hand, the average stellar mass of SFGs in the satellite halo varies, with respect to their distance to the QGs, by less than a factor of 2. This suggests that the population of SF satellites is fairly homogeneous.

Furthermore, the combined photometry in the available bands can place constraints in the shape of the SED and in the inferred FIR properties, under the  the assumption that the FIR emission of QGs is well-represented by the adopted DL07 models.  This is of particular importance, especially since none of our stacked ensembles has a formal detection (S/N$>$3) in the R-J tail ($\lambda_{\rm rest} > 350$$\,\mu$m). With these important considerations in mind we conclude that future  infrared and millimetre observations of large and representative samples at $\leq 1-2\arcsec$ resolution, are necessary to capture unequivocally the ISM mass budget and ISM conditions of high-z QGs.

\section{Conclusions}
In this work we presented a robust multi-wavelength stacking analysis of a carefully selected sample of massive QGs from the COSMOS field and in three redshift bins, $\langle z \rangle =$ 0.5, 0.9, 1.2. We further complemented our sample with the stacking 
ensemble of Gobat et al.\,(2018), at $\langle z \rangle = 1.8$,  constructed following the same selection criteria as those adopted in the current study, as well as with literature data for local QGs. By modelling the stacked photometry, we drew the characteristic FIR SEDs, inferred the ISM mass budget, and explored the evolution of the FIR properties of massive QGs over the last ten billion years. Our results are summarised as follows:  

\begin{itemize}

\item The \md/\ms\ ratio rises steeply  as a function of redshift up to $z \sim 1.0$ and then remains flat at least out to $ z = 2.0$. The evolution of \fgas\ follows (by construction) a similar trend, with a normalisation that depends on the assumed gas phase metallicity of the QGs. For solar metallicity, \fgas\ increases from 2\% to 8\% between $z = 0.5$ and $z = 1.0$.  The evolution of $f_{\text{gas}}$ in our QG samples can then be interpreted as a combination of progenitor bias at $z > 1$, a uniformity in ISM conditions among newly quenched galaxies, and a closed-box consumption of the ISM after quenching, with a depletion time of $\sim$2\,Gyr.

\item The gas depletion time scales of massive QGs at all redshifts are comparable to that of local SFGs and systematically longer than that of MS galaxies at their corresponding redshifts.   

\item The dust temperature of massive QGs remains roughly constant with \td\ = 21$\pm 2$\,K out to $z = 2.0$, with only a weak (if any) evolution towards a marginally higher \td\ at higher redshifts,  which can be  explained as the consequence of the average size evolution of QGs. This motivated us to construct and make publicly available a template IR SED of massive $z > 0$ QGs.  

\item Based on our template SED, we provide predictions for the flux densities of the continuum emission in the R-J tail of passive galaxies as a function of redshift, and highlight the need of accounting for CMB effects at $z > 3$. Finally, we argue that \mgas\ prescriptions calibrated on SFGs could lead to an overestimate of the predicted dust continuum flux density of QGs in the (sub)-mm bands and discuss the implications for ALMA observing strategies.

\end{itemize}

After accounting for progenitor bias, we are ultimately left with the picture of a weak to flat evolution of initial ISM conditions in QGs. That is, our current constraints do not require that the gas fraction of QGs immediately after quenching, nor the depletion time after that, change significantly in the 10\,Gyr between $z = 0$ and $z \sim 2$. This apparent regularity in the internal properties of QGs across most of the history of the Universe mirrors the consistency of star formation physics in MS galaxies throughout cosmic time. We conclude that this uniformity of QGs should extend to earlier $z > 2.0 $ epochs, which so far lack strong constraints in this regard.

\section*{Acknowledgements}
GEM and FV acknowledge the Villum  Fonden  research  grant  13160  “Gas to  stars,  stars  to  dust:   tracing  star  formation  across cosmic  time”  and  the  Cosmic  Dawn  Center  of  Excellence  funded  by  the  Danish  National  Research  Foundation  under  then  grant  No. 140. FV acknowledges support from the Carlsberg Foundation research grant CF18-0388 ``Galaxies:   Rise  And  Death”. S.J. acknowledges financial support from the Spanish Ministry of Science, Innovation and Universities (MICIU) under AYA2017-84061-P, co-financed by FEDER (European Regional Development Funds). K.E.W. wishes to acknowledge funding from the Alfred P. Sloan Foundation.

\bibliography{ms.bib} 
\bibliographystyle{aa}

\appendix
\section{Cutouts and satellite density}
\label{app:satellites}

\begin{figure*}[!ht]
    \centering
    \includegraphics[scale=0.65]{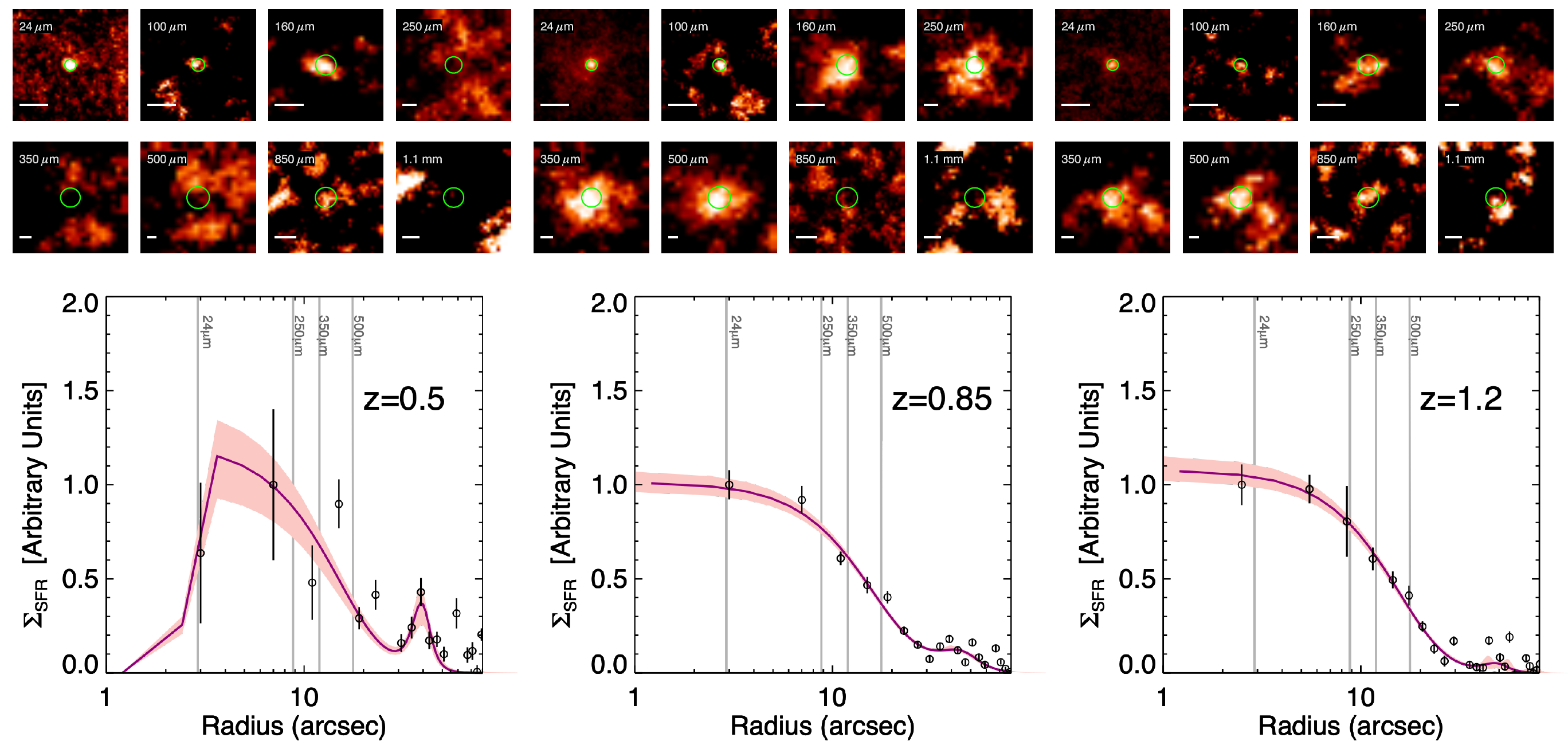}\\
    \caption{\textbf{Stacked cutouts and SFR surface density profiles}. \textit{Top:} Median stacked cutouts at the position of the QGs in the three redshift bins, at 24\,$\mu$m (\s/MIPS), 100 and160\,$\mu$m
    (\h/PACS), 250, 350 and 500\,$\mu$m \h/SPIRE, 850\,$\mu$m 
    ($JCMT$/SCUBA-2) and 1.1 mm ($ASTE$/AzTEC). The white line in each panel has a length of 15'', while the green circles show the beam FWHM in each band. \textit{Bottom:} 
    Normalised, average SFR surface density ($\Sigma_{\rm SFR}$) profile of the satellites of 
    QGs (circles) in various redshift bins. The error bars correspond to 
    the standard deviation of the surface density in each bin. The pink
    lines and the shaded regions show the best-fit `quenched' $\beta$-models to 
    $\Sigma_{\rm SFR}$ and their uncertainties. The vertical grey lines denote the  
    half-width at half-maximum of the MIPS and SPIRE beams.}

    \label{fig:images}
\end{figure*}

\begin{figure}[!h]
    \centering
    \includegraphics[scale=0.80]{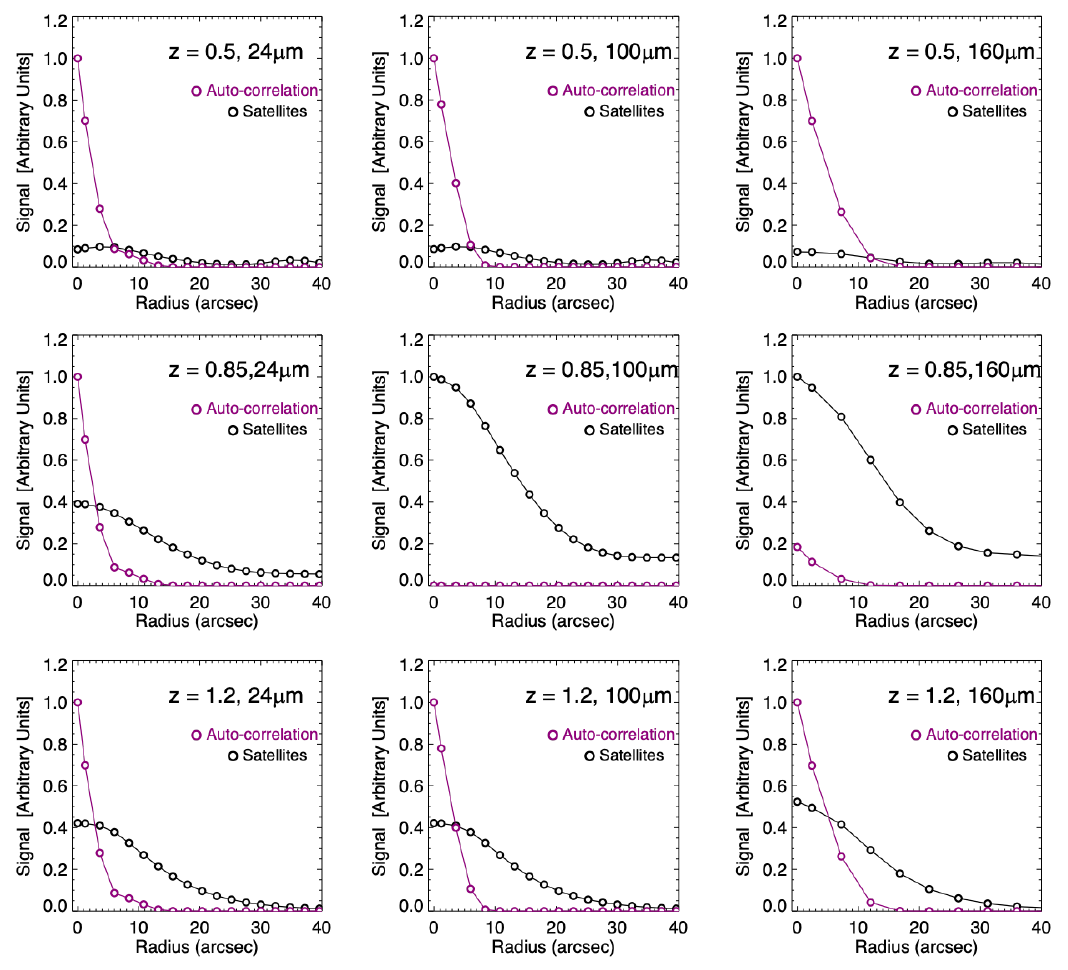}\\
    \caption{1D profiles of the auto-correlation and satellite signal, showing their respective amplitudes (in arbitrary units) at 24, 100 and 160$\,\mu$m. The autocorrelation signal includes the central point source since one fitted amplitude is used for both (see Section~\ref{sec:method}). 
    }
    \label{fig:auto}
\end{figure}

Galaxies associated with the QGs in our samples (hereafter `satellites') were selected from the \citet{laigle_2016} multi-wavelength catalogue as star forming based on the $UVJ$ criterion and with photometric redshifts compatible within their uncertainties with that of each QG. The rest-frame UV SEDs of satellites were then fitted with 100\,Myr-old composite stellar population models of constant star formation, with and without dust attenuation \citep{noll_2009}, to derive both extinction-corrected and uncorrected SFRs. We then took the difference between those two estimates as the obscured SFR of satellites, and these values were averaged in radial bins to construct for each sample (Fig.~\ref{fig:images}, bottom panels). 
The obscured SFR density distributions thus constructed were fitted with a nine-parameter combination of two $\beta$-profiles of the form $A\left[1+\left((r-r_{0})/r_{c}\right)^2\right]^{-\beta}$, the first fixed at $r_{0}=0$, and a multiplicative quenching term of the form $\min(r/r_{q},1)^{\alpha_q}$. The latter is only appreciable in the case of the low-redshift bin, as shown in Fig.~\ref{fig:images}. The two $\beta$-models describe the one- and two-halo terms of satellites,  and the latter increases noticeably in amplitude with decreasing redshift. This is not surprising as galaxies with a fixed stellar mass will be less dominant within their halo at low than at high redshift. Finally, the functional profiles were circularised and normalised to 1\,$L_{\odot}$ when fitting the stacked cutouts. 
The 2D QG (point source + autocorrelation) and circularised satellite signals are left to vary freely at 24, 100 and 160$\,\mu$m, where the instrumental beam is small enough that their different shapes can be resolved (see Fig.~\ref{fig:auto}). At longer wavelengths the satellite signal is extrapolated from the 24, 100 and 160$\,\mu$m amplitudes using \citet{magdis_2012} templates, as described in Section~\ref{sec:method}. After subtracting the central source, autocorrelation signal, and satellite profile, we find that the median residuals are $\lesssim$6\%.
\\

\begin{figure}[!h]
    \centering
    \includegraphics[scale=0.85]{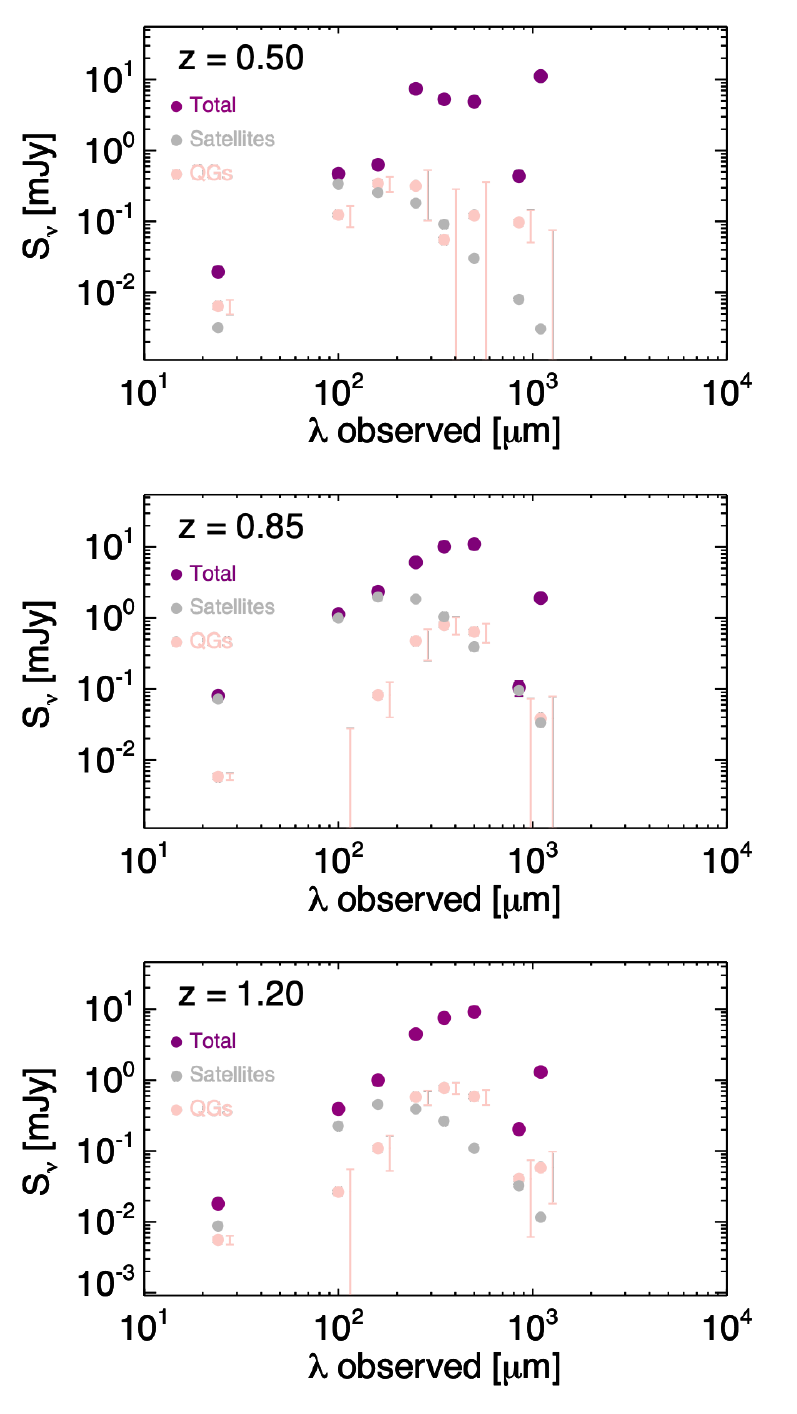}\\
    \caption{ Background-subtracted and bias-corrected FIR SEDs, before and after decomposition in SF and QG components: total signal (purple), SF satellites (grey), and QG with error bars on the side (light pink). Only the point-source component of the QG SED is shown (that is, after subtraction of the autocorrelation term). As described in Section~\ref{sec:method}, the satellite points are fitted to the data at $\leq160$\,$\mu$m and extrapolated from \citet{magdis_2012} templates at longer wavelengths.}. 
    \label{fig:decomp}
\end{figure}

This method makes the implicit assumptions that the SED-derived obscured SFR and FIR emission of satellites have the same spatial distributions, and that satellites below the detection threshold of the multi-wavelength \citet{laigle_2016} catalogue are similarly distributed as their brighter counterparts. On the other hand, we note that the median mass of satellites increases somewhat at radii $r<10''$, which suggests that sub-threshold (i.e. lower mass) satellites contribute less signal at small radii. This might result in a slight underestimate of the FIR flux of the central QG. Fig.~\ref{fig:decomp} shows, for the three redshift bins, the total measured FIR signal \citep[after background subtraction, and bias correction as in][]{bethermin_2015} and its decomposition in SF satellite and QG components (the latter after subtraction of the autocorrelation term). The values of each component in each band along with the corresponding uncertainties are listed in Table ~\ref{tab:errors}.

\begin{sidewaystable*} 
\tiny
    \centering
    
    \caption{Stack decomposition}
    \begin{tabular}{ccccccccc}
    \hline\hline 

         Redshift& $24\,\mu m$& $100\,\mu m$ & $160\,\mu m$ & $250\,\mu m$ & $350\,\mu m$& $500\,\mu m$ & $850\,\mu m$ & $1100\,\mu m$ \\
         &[$\mu Jy$] &[$mJy$]& [$mJy$]& [$mJy$]& [$mJy$]&[$mJy$] & [$mJy$]&[$mJy$]\\ 
         (1)& (2)& (3)& (4)& (5)& (6)& (7)& (8)& (9)\\ 
    \hline
        &&& & Total&&&&\\
     \hline
    $\langle 0.50 \rangle$&    $19.57\pm 0.10$&    $0.47\pm0.02$&  $0.63\pm0.07$&  $7.43\pm0.21$&  $5.30\pm0.22$&   $4.91\pm0.23$&  $0.43\pm0.04$&  $11.16\pm0.07$\\ 
    $\langle 0.85 \rangle$&   $80.07\pm 0.45$& $1.12\pm0.02 $&  $2.35\pm0.04$&  $6.08\pm0.12$&  $10.15\pm0.15$&   $10.97\pm0.18$&  $0.10\pm0.07$&  $1.91\pm0.04$\\   
    $\langle 1.20 \rangle$&   $18.06\pm 0.59$& $0.39\pm0.02$&  $0.99\pm0.04$&  $4.43\pm0.12$ &  $7.52\pm0.13$&  $9.15\pm0.14$&   $0.20\pm0.03$&  $1.30\pm0.03$\\  
    &&& & Satellites&&&&\\
     \hline
    \hline
        $\langle 0.50 \rangle$&$3.18\pm 1.00$&$0.33\pm0.03$&  $0.25\pm0.02$&  $0.18\pm0.02$&  $0.09\pm0.01$&   $0.03\pm0.01$&  $0.008\pm0.001$&  $0.003\pm0.001$\\  
       
        $\langle 0.85 \rangle$&   $72.22\pm 0.42$ & $0.99\pm0.02 $&  $1.98\pm0.01$ &  $1.85\pm0.18$ &  $1.04\pm0.15$ &   $0.39\pm0.06$&  $0.09\pm0.02$&  $0.03\pm0.01$\\   
        
        $\langle 1.20 \rangle$&   $8.78\pm 0.53$& $0.22\pm0.02$&  $0.45\pm0.03$&  $0.39\pm0.04$&  $0.02\pm0.04$&   $0.10\pm0.02$&  $0.032\pm0.01$&  $0.011\pm0.002$\\  
        
    \hline
\hline
    &&& & Background&&&&\\
     \hline
    $\langle 0.50 \rangle$&    $ 6.40e-03\pm 5.4e-05$&    $1.23e-01\pm1.4e-05$&  $ 3.24e-01\pm4.5e-05$&  $2.86e-01\pm4.5e-02$&  $3.22e-02\pm3.4e-05$&   $1.11e-01\pm 4.1e-05$&  $1.06e-01\pm1.3e-05$&  $2.66e-7\pm7.0e-05$\\ 
    $\langle 0.85 \rangle$&   $5.820e-03\pm 3.5e-05$& $0.00\pm9.6e-06 $&  $1.95e-01\pm2.9e-05$&  $5.10e-01\pm 2.2e-02$&  $1.01e+00\pm2.8e-05$&   $6.51e-01\pm1.2e-02$&  $ 2.11e-8\pm7.6e-06$&  $ 3.15e-02\pm4.2e-05$\\   
    $\langle 1.20 \rangle$&   $5.38e-04\pm 2.9e-05$& $4.29e-03\pm8.7e-06$&  $2.57e-02\pm3.3e-05$&  $ 9.09e-01\pm2.3e-02$ &  $ 4.89e-03\pm2.7e-05$&  $-2.18e-01\pm1.2e-02$&   $ 4.18e-03\pm8.3e-06$&  $-1.60e-02\pm4.6e-05$\\ 
    \hline                  
    \hline
    \end{tabular}
     
       \label{tab:errors}
\end{sidewaystable*}

\section{Robustness of the sample selection}\label{app:subsamples}
\begin{figure*}[!ht]
    \centering
    \includegraphics[scale=0.32]{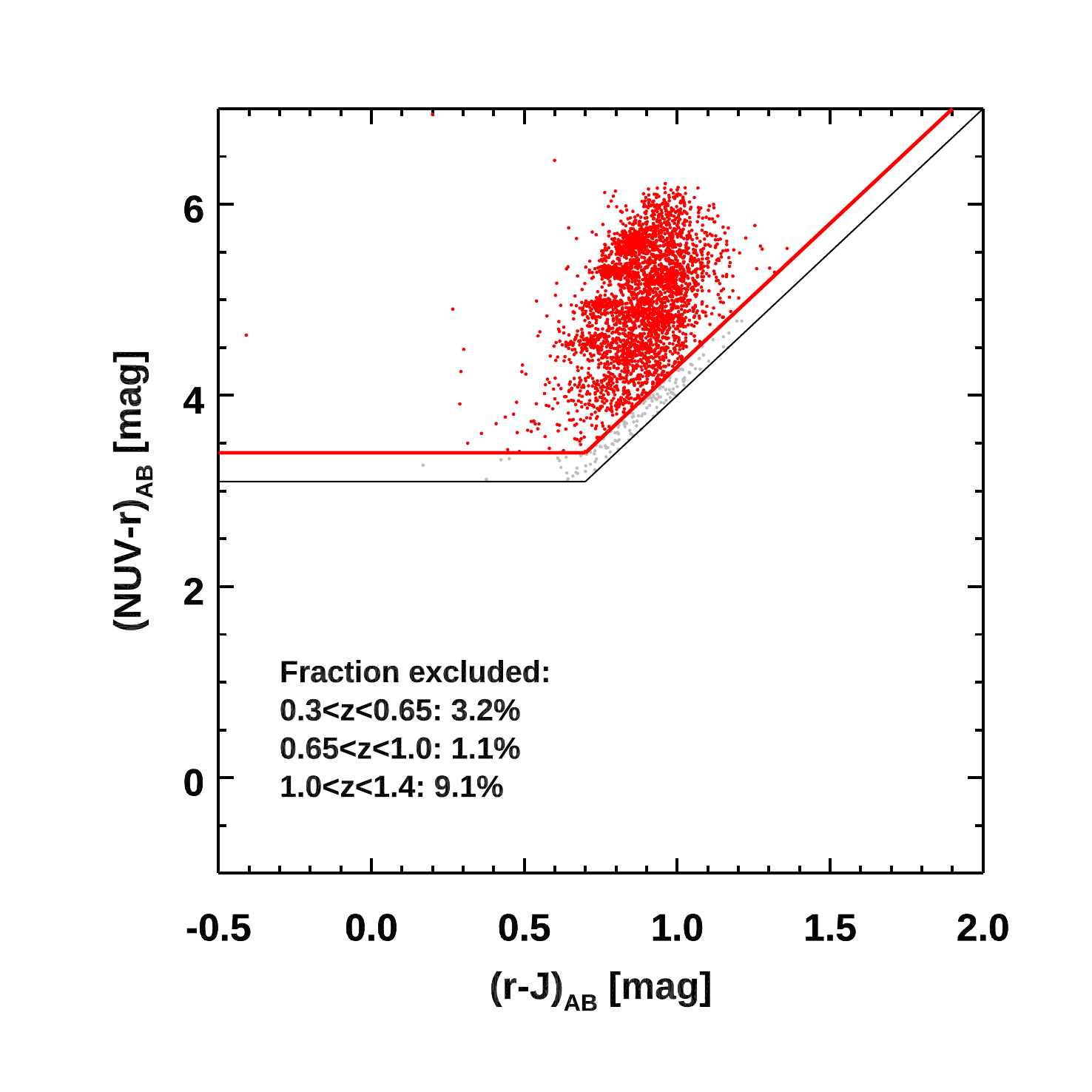}
    \includegraphics[scale=0.32]{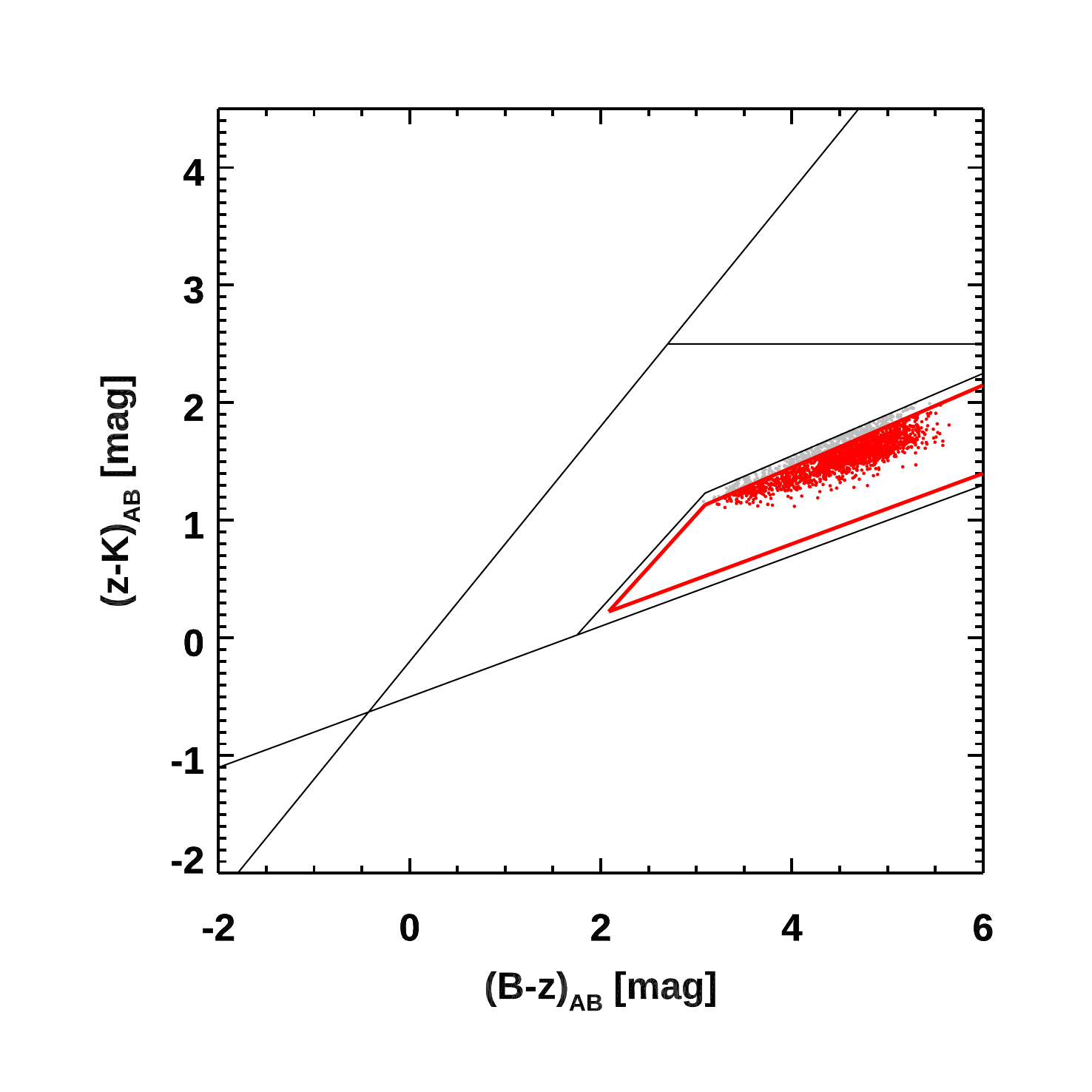}
    \includegraphics[scale=0.32]{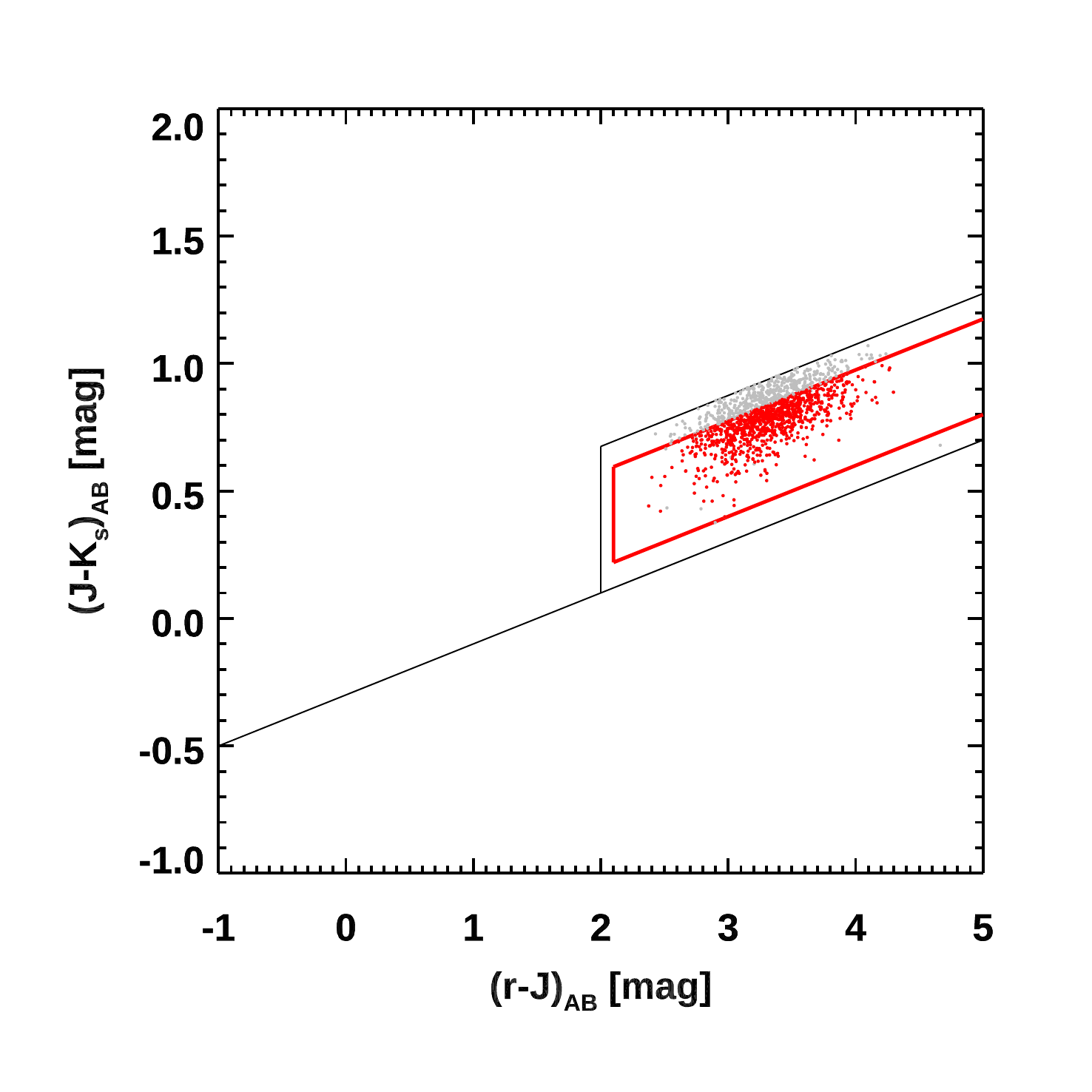}\\
    \includegraphics[scale=0.32]{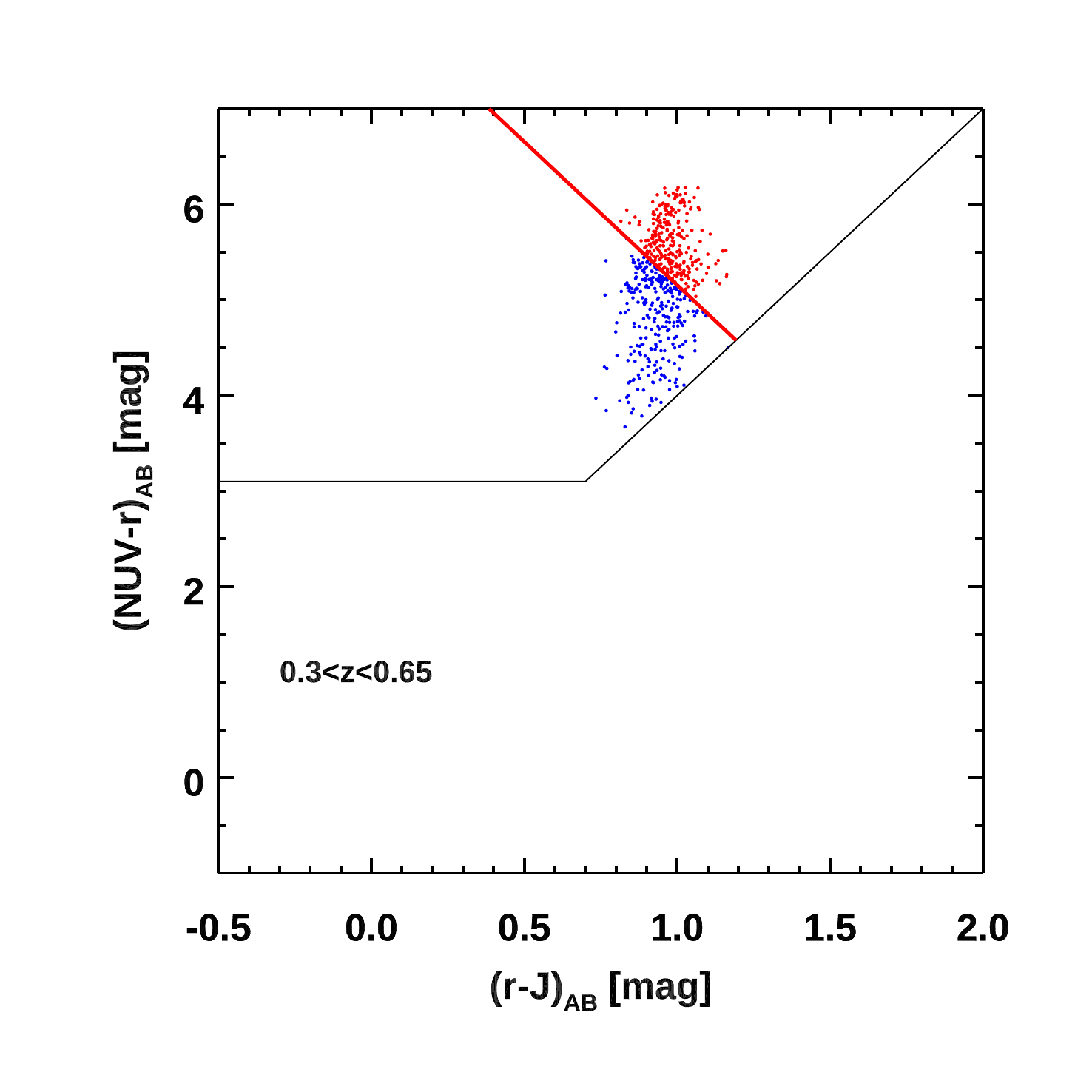}
    \includegraphics[scale=0.32]{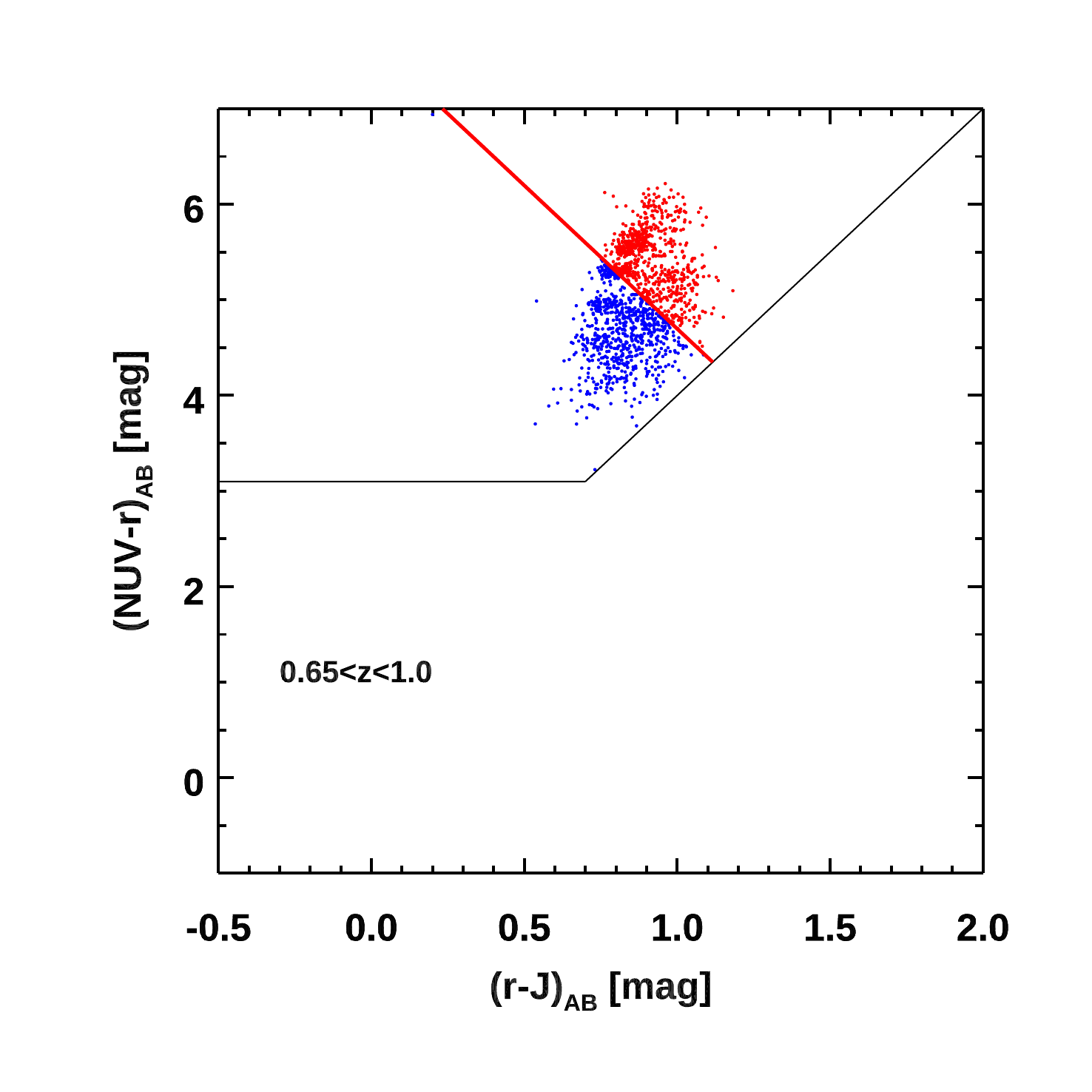}
    \includegraphics[scale=0.32]{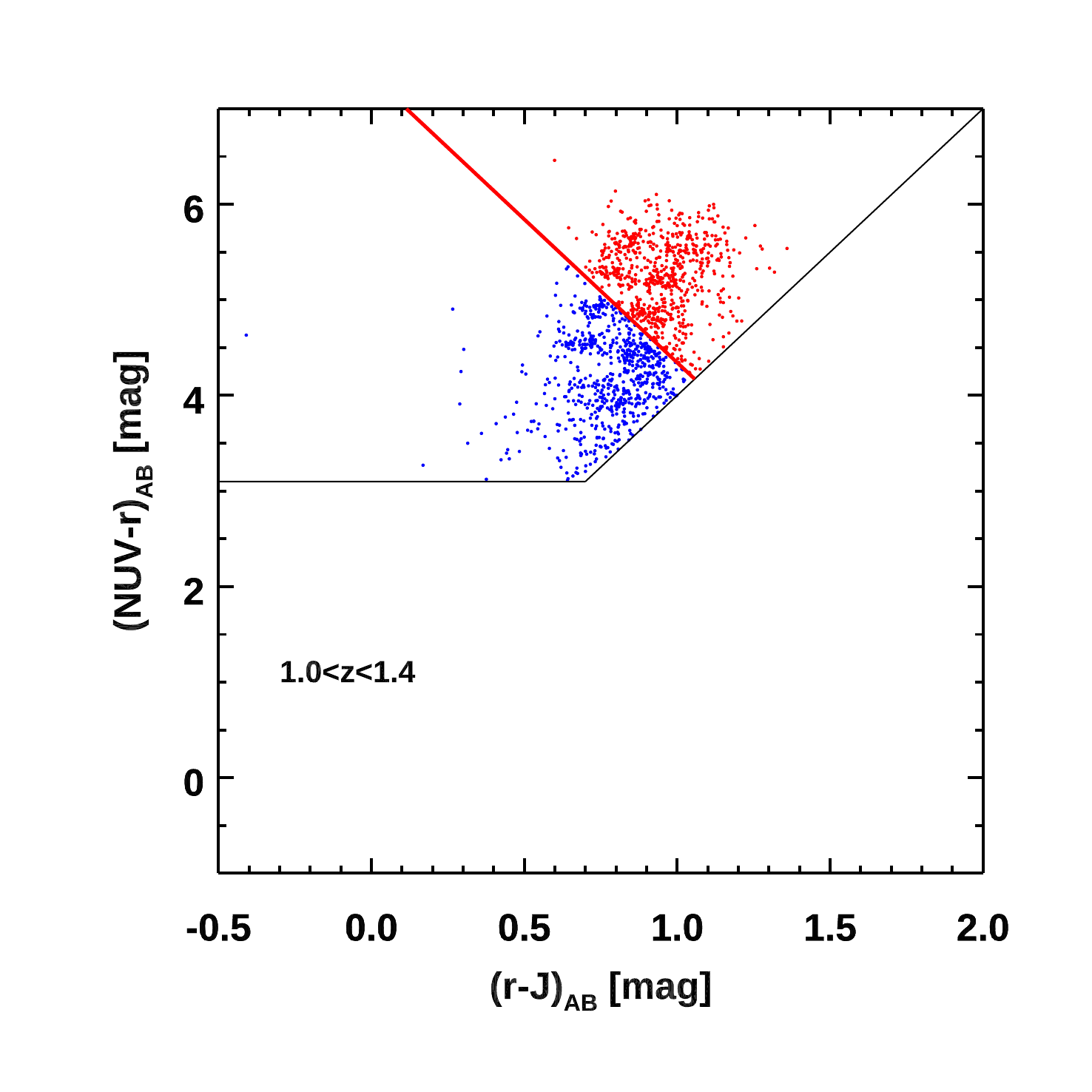}
    \caption{\textbf{Sample selection tests}. \textit{Top:} \textit{NUVrJ}, \textit{BzK}, and \textit{rJK} colour diagrams with a conservative 0.1\,mag margin (red lines) added to prevent contamination from reddened star forming objects. 
    \textit{Bottom:} low-, mid-, and high-$z$ samples divided into blue and red sub-samples in \textit{NUVrJ} colour space.}
    
    \label{fig:subsamples}
\end{figure*}

To investigate the dependency of our results on the selection criteria used to build the three samples, and in particular to conclusively exclude a non-quiescent origin for the FIR signal found, we performed the analysis on sub-samples drawn from the main one using the following a `conservative' colour cut that adds a 0.1\,mag margin to the \textit{NUVrJ}, \textit{rJK}, and \textit{BzK} quiescent loci used to select QGs (Fig.~\ref{fig:subsamples}, top). This was intended to further minimise the possibility of contamination of the sample by dusty SFGs close to the edge of the quiescent region. This yielded 909, 1241, and 415 objects in the high-, mid-, and low-$z$ redshift bins, respectively, corresponding to a reduction by 35\%, 19\%, and 26\%.  Finally, supplementary `blue' and `red' cuts, in which the quiescent loci in \textit{NUVrJ} space were split perpendicular to the dividing line so as to generate approximately equal-size sub-samples (Fig.~\ref{fig:subsamples}, bottom) we applied. 

Stacked cutouts were produced and fluxes were extracted following the method described in Section~\ref{sec:method}, while the resulting FIR SEDs were modelled as in Section~\ref{sec:modeling}. Despite the lowered S/N due to the reduced sample sizes, we found that the SEDs of the sub-samples were consistent, within their uncertainties, with those of the full samples. This is further confirmed by the parameters produced by the fit, which do not deviate appreciably with those listed in Table~\ref{tab:properties}. In particular, those derived from the blue and red sub-samples are entirely consistent with each other. Consequently, we conclude that the latter are not the product of either contamination by non-quiescent objects, or due a particular population of QGs within the main sample.

\section{Estimate of the AGN duty cycle}
\label{app:dutycyle}

After subtracting the star-forming radio slope, estimated from the FIR-radio correlation \citep{Delhaize17}, the average excess at 1.4\,GHz in the high-, mid-, and low-$z$ bins corresponds to intrinsic luminosities of $2.2\times10^{22}$, $7.0\times10^{21}$, and $5.2\times10^{21}$\,W/Hz, respectively. We also add to these the $5.4\times10^{22}$\,W/Hz 1.4\,GHz excess from the $z\sim1.8$ sample of G18. To translate these excess fluxes extracted from median stacks into an AGN fraction, we considered the evolving 1.4\,GHz AGN luminosity functions of \citet{Novak_2018}, evaluated between $10^{22}$ and $10^{28}$\,WHz. We first noted that the G18, high-, mid-, and low-$z$ excesses correspond respectively to >1, 0.59$\pm$0.9, 0.19$\pm$0.04, and 0.14$\pm$0.02 times the integrated AGN luminosity function. However, this does not account for the sampling of the redshift bins nor the non-zero noise of the original data used to make the stacks \citep[about 0.015\,$\mu$Jy/beam;][]{schinnerer_2010}, which can impact the median flux. We performed a set of 1000 simulations based on random samplings of the \citet{Novak_2018} luminosity function, with sample sizes and redshift distributions corresponding to our three redshift bins. For each simulation the duty cycle was then varied from 0 to 100\%, with `off' AGNs set to zero luminosity. Finally, we added Gaussian noise to each object based on the r.m.s. of the data, converted to intrinsic luminosities. This yielded duty cycles of $0.50^{+0.18}_{-0.20}$, $0.32^{+0.07}_{-0.08}$, $0.18^{+0.04}_{-0.05}$, and $0.23^{+0.05}_{-0.04}$ for the G18, high-$z$, mid-$z$, and low-$z$ quiescent stacks, respectively.

\section{Simulation of FIR parameters}\label{app:firsimulation}
\begin{figure*}[!ht]
    \centering
    \includegraphics[scale=0.7]{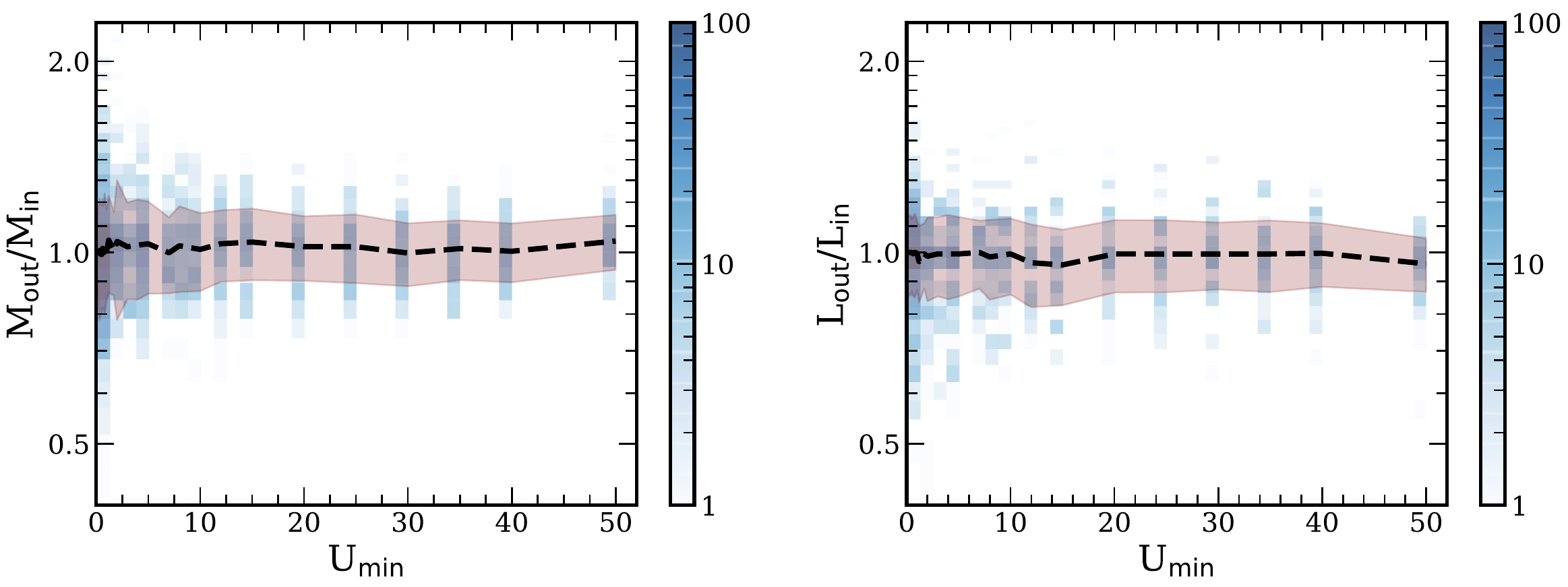}\\
    \caption{\textbf{Simulations on the derivation of \md\ and \lir}. The ratio of derived over input \md\ (left) and \lir\ (right) as a function of intrinsic $\langle U_{\rm min} \rangle$ for a range of simulated DL07 models mimicking the dynamical range and S/N of the available photometry for our stacked SEDs. For each $U_{\rm min}$ template the data are colour coded by the density of the points.  The pink shaded region encloses the 68\% of the derived ratios. Noticeably, the range of the output/input ratios, and thus the corresponding uncertainties are  larger for colder templates ($U_{\rm min} <5.0$). }
    \label{fig:sims}
\end{figure*}

To investigate the uncertainties and possible systematic effects in the derivation of the \md\ and \lir\ estimates introduced by 
the available photometry of our stacked SEDs as a function of 
the intrinsic shape of the SED, we performed bootstrapping 
simulations. We first built the full set of DL07 models ranging 
from $U_{\rm min} = 0.1$ to $50$ with a fixed \lir\ and \md. We 
then redshifted each template to the central redshift of each of 
our redshift bins and measured its synthetic photometry in all 
the available bands covered by our stacking analysis.  To mimic 
our observations from the emerging set of 8 photometric data 
points (24 $\mu$m - 1.1 mm), we added a random fluctuation in each band of up to 20\% and randomly selected and placed at S/N = 3 four data 
points between $24-1100\,\mu$m. In the process, we ensure that 
the data set included one data point at observed 24\,$\mu$m and 
one data point in the R-J tail, but not longer than $\lambda_{\rm rest} \approx 220$ $\mu$m, as is the case for all of our observed 
stacks. The remaining bands were placed at S/N = 1 and treated as upper limits.  Following this 
procedure, we constructed 100 sets of synthetic photometry for each $U_{\rm min}$ and performed SED fitting using the same methodology as that applied to our observations. The comparison between in 
output and input \md\ and \lir\ for each template is presented 
in Figure~\ref{fig:sims}. It is evident that the uncertainties 
become more severe for the coldest templates. This is expected 
since for the coldest templates with the lowest $U_{\rm min}$ 
the $\lambda^{\rm max}_{\rm rest} \approx 220\,\mu$m is moving 
closer to the peak of the SED, leaving the R-J tail less 
constrained, with implications on the derived \md. Nevertheless, we see that for the combination of the data points considered 
here, there are no systematic effects in the derivation of \md\ 
and \lir. 

\end{document}